\documentclass[aps,prb,reprint,twocolumn,showpacs,superscriptaddress]{revtex4-1}
\usepackage{amsmath, amssymb, braket, dsfont, times, units, ulem}
\usepackage[mathscr]{euscript}
\usepackage{graphicx}
\usepackage{caption}
\usepackage{subcaption}

\usepackage[breaklinks,colorlinks,citecolor=magenta]{hyperref}			
\usepackage{mathtools}

\DeclarePairedDelimiter\floor{\lfloor}{\rfloor}
\def \n{{\mathbf n}}
\def \m{{\mathbf m}}
\def \r{{\mathbf r}}

\usepackage{xcolor}

\begin{document}
\title{Skyrmion Superconductivity: DMRG evidence for a topological route to superconductivity}
\author{Shubhayu Chatterjee}
\affiliation{Department of Physics, University of California, Berkeley, CA 94720, USA}
\author{Matteo Ippoliti}
\affiliation{Department of Physics, Stanford University, Stanford, CA 94305, USA}
\author{Michael P. Zaletel}
\affiliation{Department of Physics, University of California, Berkeley, CA 94720, USA}
	\affiliation{Materials Sciences Division, Lawrence Berkeley National Laboratory, Berkeley, California 94720, USA
}
\date{\today}							

\begin{abstract}
It was recently suggested that the topology of magic-angle twisted bilayer graphene's (MATBG) flat bands could provide a novel mechanism for superconductivity distinct  from both weakly-coupled BCS theory and the $d$-wave phenomenology of the high-$T_c$  cuprates.
In this work, we examine this possibility using a density matrix renormalization group (DMRG) study of a  model which captures the essential features of MATBG's  symmetry and topology.
Using large scale cylinder-DMRG calculations to obtain the ground state and its excitations as a function of the electron doping, we find clear evidence for superconductivity driven by the binding of electrons into charge-$2e$ skyrmions. 
Remarkably, this binding is observed even in the regime where the unscreened Coulomb repulsion is by-far the largest energy scale, demonstrating the robustness of this topological, all-electronic pairing mechanism.
\end{abstract}

\maketitle

A prerequisite for superconductivity is the binding of charge-$e$ fermions into bosonic charge-$2e$ Cooper pairs \cite{BCS}. This requires an attractive  interaction between two fermions which carry the same charge, and consequently must overcome their natural tendency to stay apart due to Coulomb repulsion. Conventional lore dictates that these charge carriers are electrons, and that the attraction is mediated by low-energy bosonic collective modes of lattice (phonons) or electronic (critical fluctuations or Goldstone modes) origin \cite{BCS,KohnLuttinger,ScalapinoReview,MaitiChubukov}. However, quantum materials with topologically non-trivial band structures can intertwine spin and charge degrees of freedom,  leading to solitonic spin-textures called \emph{skyrmions}\cite{Skyrme,BP75} which carry electrical charge \cite{Sondhi,MoonMori}. This naturally begs the question: Can superconductivity arise from pairing of charge-$e$ skyrmions, rather than electrons? And what might provide the ``pairing glue'' between skyrmions that enables them to overcome Coulomb repulsion? 

In a companion work\cite{Eslam}, we analytically argued that magic angle twisted bilayer graphene (MATBG) has the requisite band topology and symmetries to exhibit superconductivity via skyrmion-pairing \cite{AbanovWeigmann,GroverSenthil,Assaad,Chakravarty}. 
Recent experimental evidence in favor of strong coupling superconductivity\cite{Tian2022} and the presence of charged skyrmions in MATBG\cite{Yu2022} further motivates a thorough, numerically unbiased investigation of skyrmion-pairing.
In this work, we distill the essential features of MATBG into a minimal model for skyrmion superconductivity which we explore using  large-scale density matrix renormalization group (DMRG) \cite{white} calculations. We find concrete numerical evidence for a skyrmion-pairing mechanism that requires neither retardation nor screening. Our work thus confirms the viability of a novel strong-coupling route to superconductivity which is all-electronic in nature, providing a new avenue in the search for superconductivity at higher temperatures. 

\begin{figure}
    \centering
    \includegraphics[width=0.4\textwidth]{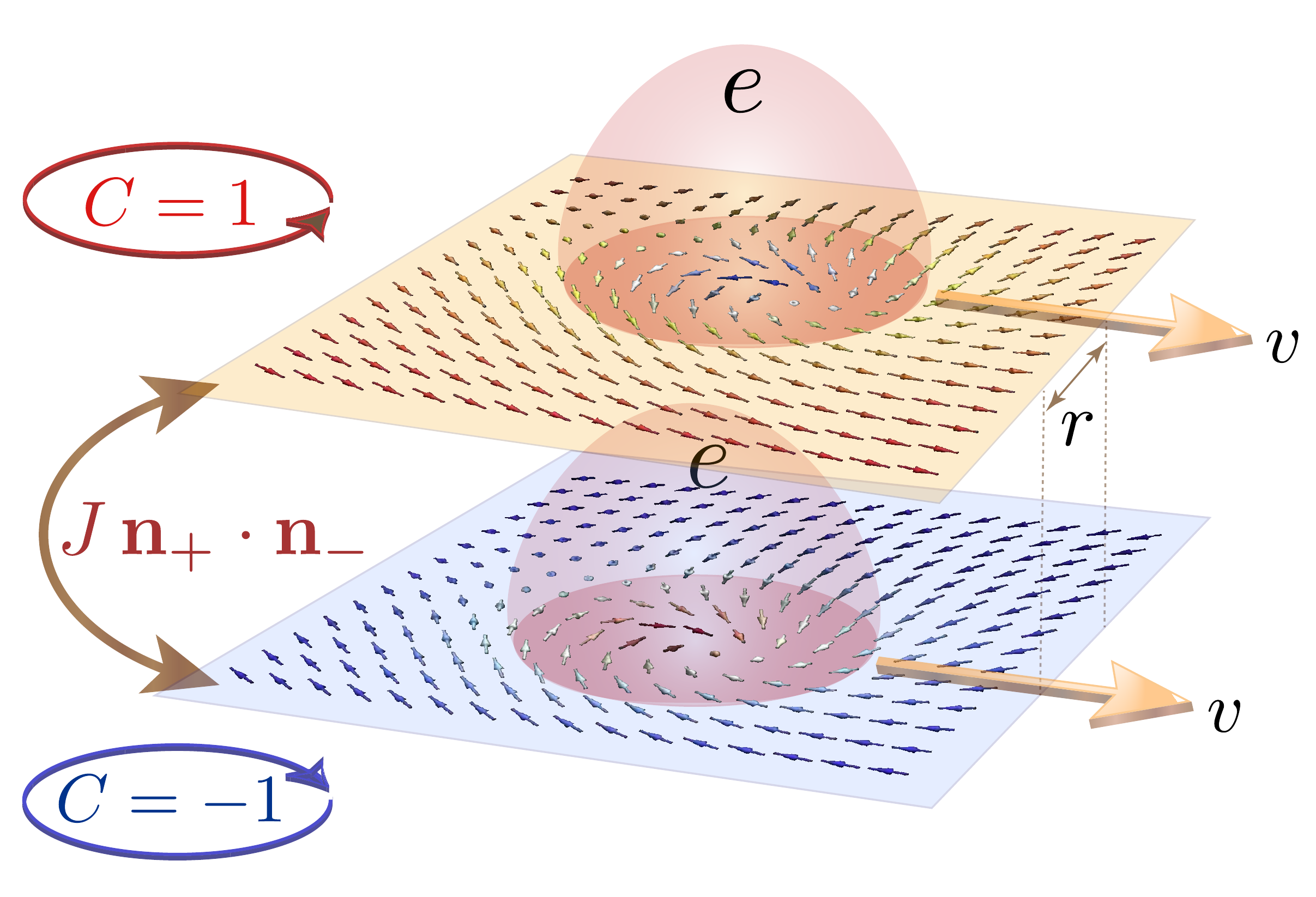}
    \caption{Schematic depiction of mobile charge-$2e$ skyrmion-pair excitation over an easy plane antiferromagnetic ground state in a bilayer with opposite magnetic fields. Pairing of charge-$e$ skyrmions in opposite layers is induced by local antiferromagnetic exchange $J$, which is sufficient to overcome the long-range Coulomb repulsion because of the large spatial spread of the skyrmions.}
    \label{fig:SkPairCartoon}
\end{figure}

To seek out the basic ingredients for this physics, it is useful to recount some essential features of MATBG.
MATBG features eight flat bands arising from spin, valley, and an additional orbital degree of freedom ``$\gamma = \pm$.''
Crucially, in the basis where the orbital index transforms naturally under the space-group symmetries,
the four $\gamma = +$ bands have Chern number $C=1$, while the four $\gamma = -$ bands have $C=-1$.
Appealing to the equivalence between Chern bands and the quantum Hall effect, MATBG can thus be viewed as a bilayer of $U(4)$ quantum Hall systems, but with opposite layers seeing opposite magnetic fields (Fig. \ref{fig:SkPairCartoon}) \cite{Tarnopolsky,XiDai,NickPRX,Eslam,VafekKang}.

By analogy to quantum Hall ferromagnetism, at integer fillings the electrons may spontaneously polarize along  axes of the spin-valley-orbital space and form insulators. 
Small terms in the Hamiltonian which break the approximate symmetry down to the exact symmetries of  charge, valley, and spin, $U(4) \times U(4) \to U_C(1) \times U_V(1) \times SU_{S}(2)$ determine the precise nature of the symmetry breaking. 
Regardless of these details, the enlarged approximate symmetry leaves behind a signature: soft bosonic modes coming from fluctuations in the $U(4) \times U(4)$ space which are described by a non-linear sigma model (NL$\sigma$M) with topological terms \cite{NickPRX,Eslam,VafekKang,KBVZ}.

When MATBG is doped away from certain integer fillings, superconductivity is observed \cite{PabloMott,PabloSC,Dean-Young,YoungScreening,efetov,EfetovScreening,Arora}.
Superconductivity requires two ingredients:  a pairing mechanism, and  a superfluid stiffness $\rho_{\textrm{SC}}$ to establish phase coherence.
Several works have recently emphasized how the topology of the MATBG flat bands  might enhance  $\rho_{\textrm{SC}}$.\cite{Andrei, Torma, Rossi,peri2020fragile}
However, this effect doesn't provide a reason for electrons to pair in the first place. 
In Ref.~\onlinecite{Eslam} it is argued that the topology of MATBG  may play a crucial role in the pairing mechanism as well (see also Ref.~\onlinecite{Christos}). 

The NL$\sigma$M describing fluctuations in the $U(4) \times U(4)$ pseudo-spin space
admits topological textures, skyrmions, which carry charge-$2e$. It was argued that the charge-$2e$ skyrmion is stable against disassociating into two charge-$1e$ electrons
\emph{even in the presence of a long-range, unscreened} Coulomb interaction, providing an all-electronic pairing mechanism.\cite{Eslam}
When the system is doped away from electron filling $\nu = 0, \pm 2$, the charge enters in the form of these bosonic skyrmions, which may Bose-condense and lead to superconductivity. 

\textit{The model.}
In this work we numerically investigate this proposal in a phenomenological model where the Chern bands of MATBG are instead modelled as Landau levels. 
The physical electron spin, while important for understanding the full MATBG phase diagram, is not essential for the pairing mechanism, so here we neglect it and work with a spinless four-component model.
There are a variety of scenarios for how this spinless model  embeds into the MATBG phase diagram,\cite{Eslam} but as one concrete example, MATBG may be spin polarized in the vicinity of $\nu=-2$, in which case our model describes the half-occupied spin species. 
The four remaining bands are labeled by a  ``layer'' index $\gamma^z$ and a ``isospin'' index $\eta^z$.
The precise relation between $\gamma, \eta$ and the MATBG degrees of freedom is not so important,\cite{Eslam} but we note that $\eta^z$ is in fact the valley index.

The two essential ingredients for skyrmion superconductivity  are that (1) the bands carry Chern number $C = \gamma^z = \pm 1$ and (2) there is an anti-ferromagnetic interaction between the isospin of the two layers in addition to the long-range Coulomb repulsion. 
In terms of the electron field operators $\psi_{\gamma \eta}(r)$, we thus consider the following 2D continuum model:
\begin{align}
H &=  \psi^\dagger \frac{(\mathbf{p} + e \gamma^z \mathbf{A})^2}{2m} \psi   + \frac{1}{2} \int : n(r) V_C(r - r') n(r') :  \notag \\
 & \quad    -  E_C \ell_B^2 \sum_{i=x, y, z} J_i   : \left( \psi^\dagger \gamma^z \eta^i \psi(r) \right)^2 :
 \label{eq:H}
\end{align}
The layers see \emph{opposite} magnetic field $\nabla \times \mathbf{A} = B$. 
Here $V_C(r)$ is the Coulomb repulsion,  $n(r) = \sum_{\gamma \eta} \psi^\dagger_{\gamma \eta} \psi_{\gamma \eta}(r)$ is the charge density, and  $J_x=J_y=J + \lambda$, $J_z = J - \lambda$ parameterize an anti-ferromagnetic XXZ interaction between the two layers.
We account for proximate metallic gates at distance $d$ by taking
$V_C(\mathbf{q}) = \frac{2 \pi}{q} \tanh(q d)$, expressed in units of the magnetic length $\ell^2_B = \hbar / e B$ and Coulomb energy $E_C =\frac{e^2}{4 \pi \epsilon \ell_B }$.
We fix $d = 3 \ell_B$ to match typical gate distances in MATBG devices under the identification $2 \pi \ell^2_B = A_M$,  where $A_M$ is the area of moir\'e cell. 

We note that in the context of MATBG, $J$ arises when treating the flat-band dispersion within second-order perturbation theory.\cite{NickPRX, Eslam, LianIV} In the present model, this dispersion corresponds to a small tunnel coupling between the two layers, $t \psi^\dagger \gamma^x \psi$, which can be treated perturbatively near filling $\nu = 2$ to obtain $J \propto t^2 / E_C$, in close analogy to super-exchange.\cite{dassarma98, NickPRX} It is thus a generic feature of tunnel-coupled Chern bands. 

The Hamiltonian Eq.~\eqref{eq:H} is then projected into the lowest Landau level (LLL) of each component, quenching the kinetic energy.
We note at the outset that the resulting model is \emph{entirely} repulsive. 
Naively, it may look like the $J$-term puts in attraction ``by hand,'' since anti-ferromagnetically aligned electrons see a short-distance attractive interaction $V_{+\rightarrow, -\leftarrow}(r) = V_C(r) - 2 (J + \lambda) E_C \ell_B^2 \delta^{(2)}(r)$.
However, this interaction is smeared-out over the scale $\ell_B$ due to Landau level projection, and we have verified (App.~\ref{app:bare_repulsion}) that for $d = 3 \ell_B$ the projected interaction  is repulsive in all channels for $J + \lambda < 3.25$, while we work exclusively in the regime $J, \lambda \leq 1$.
So superconductivity in this model  requires an all-electronic pairing mechanism for overcoming the Coulomb repulsion.

The symmetries of the model play an important role in our analysis.  When $J, \lambda = 0$, the model is symmetric under $U(2) \times U(2)$ transformations within each layer,  the spinless analog of MATBG's $U(4) \times U(4)$.
Setting $J \neq 0, \lambda = 0$ breaks this symmetry down to $U(1) \times U(1) \times SU(2)$, which is the spinless analog of  MATBG in the ``chiral limit.'' \cite{Tarnopolsky,NickPRX}
Finally, the easy-plane anisotropy $\lambda$ further reduces the symmetry to  $U(1)^3$, corresponding to electron charge, layer polarization, and isospin $\eta^z$, (in MATBG,  the valley-U(1) symmetry).
The model also has time-reversal symmetry, $\mathcal{T} =  \gamma^x \eta^x K$, as well as a ``Kramers'' time-reversal $\mathcal{T}' =  i \gamma^x \eta^y K$, with $(\mathcal{T}')^2 = -1$.

\begin{figure}[t]
    \includegraphics[width=0.95\columnwidth]{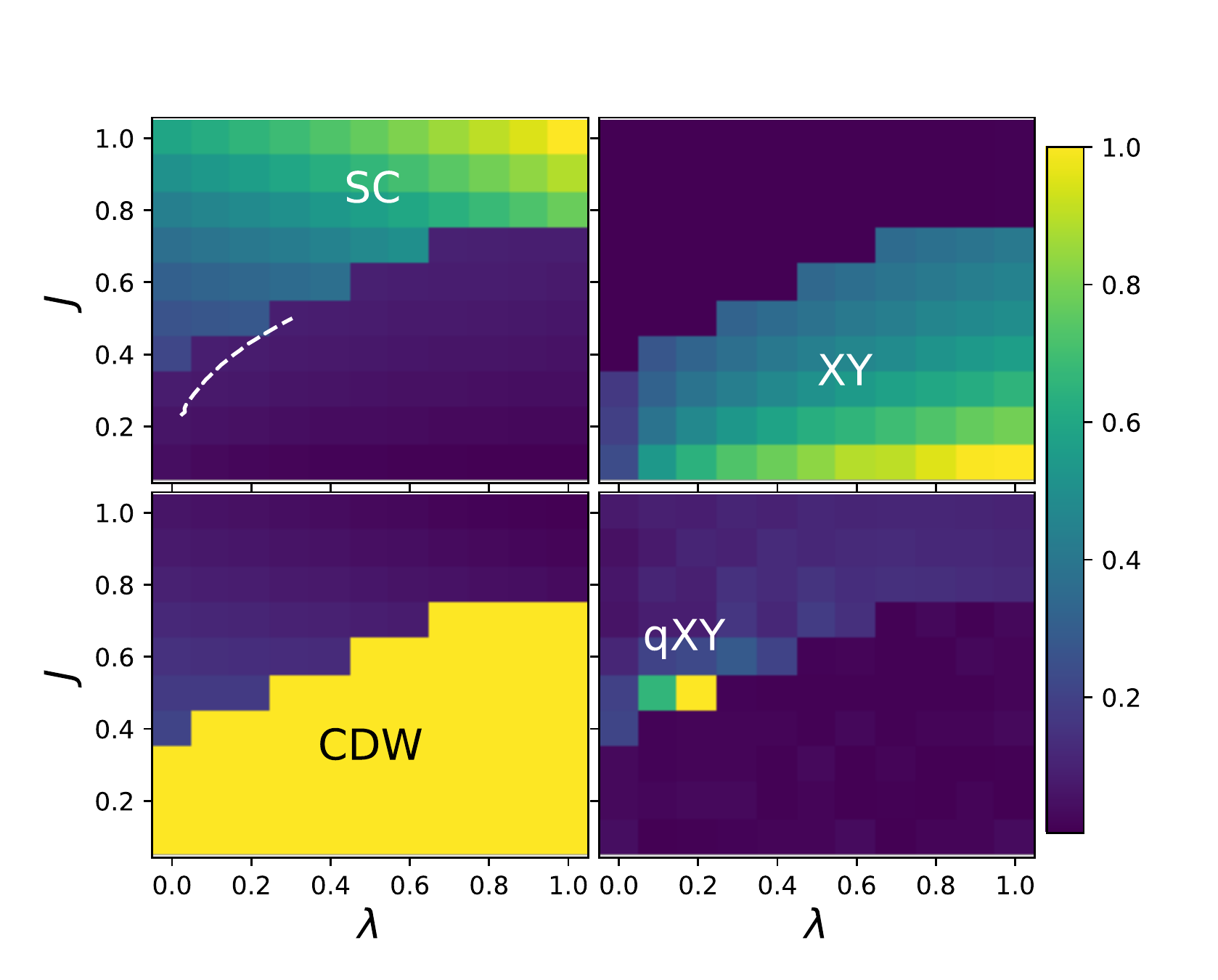}
    \caption{Phase diagram at density $\nu = 2 + \frac{1}{4}$ in the plane of $J, \lambda$, calculated at $L_y = 10 \ell_B$. Four color plots show the strength (in arbitrary units) of: superconductivity (SC), XY magnetism (XY), charge density wave order (CDW), and XY magnetism at finite wavevector (qXY). 
    There are two phases: for large-$J$ the state is a SC, while for small-$J$ the $\delta = 1/4$ doped electrons  polarize onto one ``layer'' and form a CDW which coexists with the XY order found at $\nu = 2$.
    The dashed-white line in the SC panel shows the critical $J_\ast$ for which electrons bind in to charge-2e skyrmions, as analyzed in Fig.~\ref{fig:Excitations}b.
    In a small region of the SC near the phase boundary, the SC coexists with  qXY order.
For a precise definition of the quantities shown here, see App.~\ref{App:iDMRG}, and for the phase boundary for other $\nu = 2 + \delta$, see App.~\ref{App:doping}}\label{fig:phase_diagram}
\end{figure}

\begin{figure*}[t]
    \includegraphics[width =\textwidth]{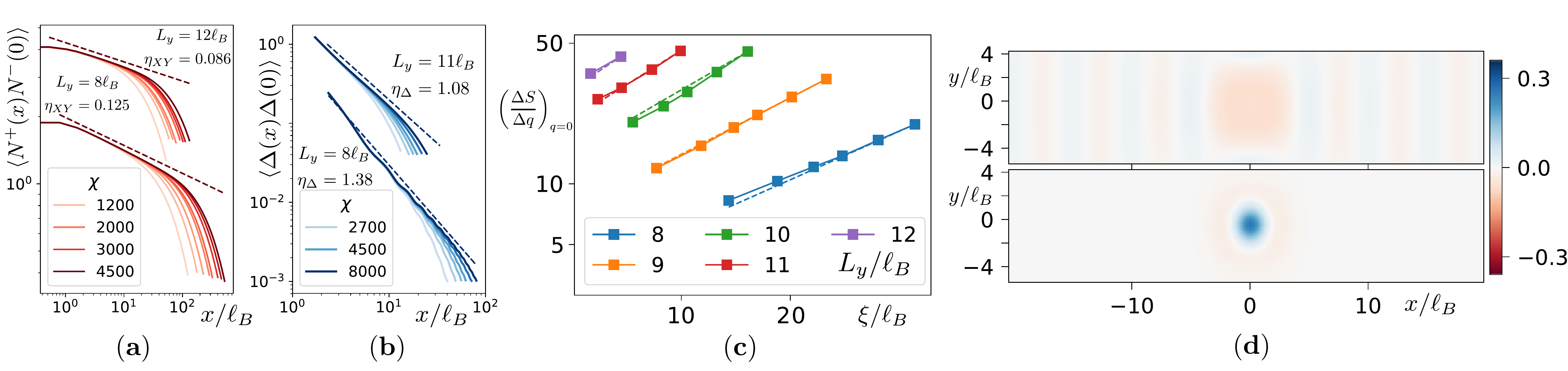}
    \caption{ \textbf{(a}) XY correlation function at $\nu = 2$, $J = 0.5, \lambda = 0.2$. In order to focus on the dependence along the length of the cylinder, $\langle N^+(x) N^-(0)\rangle$, the fields are averaged around the cylinder, $\vec{N}(x) \equiv L_y^{-1} \int dy \vec{N}(x, y)$. 
    Data is shown for two circumferences, $L_y = 8,12 \ell_B$, with the curves shifted vertically by an arbitrary displacement for clarity. 
    For each $L_y$, we show the convergence of the correlations with the MPS bond dimension $\chi$; as $\chi \to \infty$, the curves converge to a power-law with an exponent $\eta_{\textrm{XY}} \propto L^{-1}_y$ \textbf{(b}) Analogous plot for the SC order parameter $\Delta$ (Eq.~\eqref{eq:DeltaSC}) at $\nu = 2 + \frac{1}{4}$, $J = 0.9$, $\lambda = 0.6$. We again find an exponent $\eta_{\textrm{SC}} \propto L^{-1}_y$, consistent with true long-range SC order in the 2D limit. 
    \textbf{(c}) Quantitative demonstration of the relation $\eta_{\textrm{SC}} \propto L^{-1}_y$ via finite size and entanglement scaling. Here $\xi$ is the correlation length of the DMRG ground state induced by the finite-$\chi$ MPS, and $\Delta q \equiv \xi^{-1}$. $S(q)$ is the Fourier transform of the SC-SC correlation function, from which we form the scaling function $(S(0) - S(\Delta q)) / \Delta q \propto \xi^{2 - \eta_{\textrm{SC}}}$. 
    Including a range of $\xi(\chi)$ and $L_y$ for a point deep in the SC, the data is well fit by a single ansatz $\eta_{\textrm{SC}} = 9.6 / L_y$.
    \textbf{(d}) Density-density pair correlation function $g(\mathbf{r}) = (2 \pi \ell_B)^2 \langle (n(\mathbf{r}) - \nu) (n(0) - \nu) \rangle$ in the CDW phase (top, $\nu = 2 + \frac{1}{4}, J = 0.3, \lambda = 0.6$ ) and SC phase (bottom, $\nu = 2 + \frac{1}{4}, J = 0.9, \lambda = 0.6$). 
    The CDW shows long-range order, while the SC shows a short-range attractive correlation.
    }
    \label{fig:correlations}
\end{figure*}
Landau level quantization leads to a finite density of states (one per component and flux quantum), making this model amenable to numerical study much like  usual fractional quantum Hall systems. 
Here we study the model using iDMRG to obtain the ground state of Eq.~\eqref{eq:H} on an \emph{infinitely} long cylinder of circumference $L_y \sim 8 - 12 \ell_B$, where the opposite magnetic fields can be treated using a small modification of our existing QH-DMRG algorithms\cite{zaletelmulticomponent, ippolitiSO5}.
The accuracy of the DMRG is controlled by the ``bond-dimension'' $\chi$ of the associated matrix product state  ansatz, with an exact result for the ground state recovered in the limit $\chi \to \infty$.

\textit{The $\nu=2$ ``correlated insulator.''} The phase diagram of Eq.~\eqref{eq:H} depends on the density, $n = \frac{\nu}{2 \pi \ell_B^2}$ and the magnetic interactions $J, \lambda$.
The filling runs from $0 < \nu < 4$, where $\nu = 2$ is analogous to the neutrality point of spinless MATBG.
Let us first consider the state we expect to find at $\nu = 2$, where half  the LLs are filled.
By analogy to a quantum Hall ferromagnet, the Coulomb interaction will prefer to polarize the system into a spatially uniform occupation of two of the four  components, leading to a charge insulator which spontaneously breaks the isospin symmetry.
There are many ways to do so, but the anti-ferromagnetic interaction $J$ prefers  for electrons to distribute evenly between the two layers with equal and opposite isospin, so that 
\begin{align}
\vec{N}(\mathbf{r}) = 2 \pi \ell_B^2\langle \psi^\dagger(\mathbf{r}) \gamma^z \vec{\eta} \psi(\mathbf{r}) \rangle
\end{align}
orders. For example, the electrons may completely fill the $\ket{+, \rightarrow}, \ket{-, \leftarrow}$ LLs. 
This order occurs for any strength $J>0$ by consideration of the Stoner criterion: the bands are flat, so the density of states is infinite, while polarizing the electrons gains a large exchange energy of order $E_C$.
The $\lambda > 0$ anisotropy  prefers order in the XY-plane, $N^{\pm} = N^x \pm i N^y = |N| e^{\pm i \theta_{\textrm{XY}}}$. 
The XY-order spontaneously breaks $\eta^z$-rotations and time-reversal $\mathcal{T}$, while preserving the Kramers $\mathcal{T}'$, making it the analog of the ``Kramers intervalley coherent state'' identified as the ground state of MATBG at even filling in Ref.~\onlinecite{NickPRX}. 

Using iDMRG simulations  to find the ground state at $\nu = 2$, we indeed find a charge-insulator with XY-order, with one caveat. Because we consider an infinitely long cylinder,  the
Mermin-Wagner theorem implies $N^{x/y}$ can only order algebraically along the cylinder.
Consequently we  find the XY-correlations along the cylinder decay as $\langle N^{+}(x, 0) N^{-}(0, 0) \rangle \propto x^{- \eta_{\textrm{XY}}}$ with an exponent $\eta_{\textrm{XY}} \ll 1$, as shown in Fig.~\ref{fig:correlations}a.
Comparing different circumferences, we find that the exponent $\eta_{\textrm{XY}}$ decreases as $L_y^{-1}$, consistent with the transition to true long-range order in 2D. 
Using the excited state DMRG energies  we will subsequently discuss, we  find that this state has a charge gap of order $E_C$ (e.g. $\Delta_{\textrm{PH}} = 2.05 E_C$ at $J = 0.4, \lambda = 0.4$.)

\textit{The doped phase diagram.}
We then dope to density $\nu = 2 + \delta$.
A-priori the extra charge may prefer to either  distribute evenly  between the two layers, $\nu_{\pm} = 1+  \delta/2$ (``layer unpolarized''), or to polarize onto one layer, $\nu_+ = 1 + \delta, \nu_- = 1$ (``layer polarized''), so we are careful to numerically check the preferred  polarization at each point in the phase diagram.

For dopings $-1 \leq \delta \leq 1$, we find two phases in the $(J, \lambda)$-plane (Fig.~\ref{fig:phase_diagram}): for large-$J$ a layer-unpolarized superconductor (SC), and for small-$J$ a layer-polarized state which coexists on top of the XY-order. 
The precise nature of the low-$J$ state depends on the doping $\delta$, so for concreteness we discuss $\delta = \frac{1}{4}$ and refer to App.~\ref{App:doping} for other dopings. 
In this case, we find a charge density wave (CDW) order in which density $\delta$ of the electrons form a Wigner crystal in one layer, which can be detected either by a modulation in the density $\langle n(x) \rangle$ along the cylinder or by inspection of the pair-correlation function $g(\mathbf{r}) = \langle n(\mathbf{r}) n(0) \rangle$ (Fig.~\ref{fig:correlations}d).
The CDW order coexists on top of the same XY-order found at $\nu=2$.
Above a critical $J > J_c(\lambda, \delta, L_y)$, there is a first-order phase transition at which the CDW disappears and layer-unpolarized algebraic superconductivity emerges. 

The SC is an isospin singlet, pairing electrons related by the $\mathcal{T}'$ Kramers time-reversal symmetry: 
\begin{align}
\Delta(\mathbf{r}) = i \eta^y_{ij} \psi^\dagger_{+, i}(\mathbf{r}) \psi^\dagger_{-,j}(\mathbf{r})\label{eq:DeltaSC}
\end{align}
Similar to  the XY-order, we find (see Fig.~\ref{fig:correlations}b) that $\langle \Delta^\dagger(x, 0) \Delta(0, 0) \rangle \propto x^{- \eta_{\textrm{SC}}}$ with an exponent $\eta_{\textrm{SC}}(J, \lambda, \delta, L_y)$ that varies as a function of the parameters \cite{Devereaux}.
The pair carries zero \emph{orbital} angular momentum, so is in this sense analogous to an $s$-wave SC, and in App.~\ref{app:pairwf}  we further rule out pairing in higher-angular momentum channels.
However,  the  $s$-wave nomenclature is not necessarily appropriate for the potential realization in MATBG, where the electron spin (neglected here) may be polarized.

To verify that the algebraic SC converges to true off-diagonal long-range order in the 2D limit, we fix a point  in the SC region and examine the scaling of $\eta_{\textrm{SC}}$ with $L_y$, Fig.\ref{fig:correlations}c.
The scaling is consistent with $\eta_{\textrm{SC}} \propto L_y^{-1}$, which we can understand as follows.
In 2D, the SC phase fluctuations come at energy cost $E_{\textrm{2D}} = \frac{\rho_\textrm{SC}}{2} \int dx dy  (\nabla \phi)^2$, where $\rho_\textrm{SC}$ is the superfluid stiffness.
When the 2D model is placed on a cylinder, the fluctuations around the cylinder become gapped, so we can integrate $y$ to obtain $E_{\textrm{1D}} = L_y \frac{\rho_\textrm{SC}}{2} \int dx  (\partial_x \phi)^2$.
The effective 1D stiffness $L_y \rho_\textrm{SC}$ then determines the exponent $\eta_{\textrm{SC}} = \frac{2 \hbar v}{L_y \rho_{\textrm{SC}}}$, where $v$ is a velocity, as observed.
This implies $\rho_{\textrm{SC}}$ is finite as $L_y \to \infty$.

While the electron pair $\Delta(\mathbf{r})$ is gapless, we find  that all single-electron excitations are  gapped.
As a first check, we observe that the electron correlation function $\langle \psi^\dagger_i(\mathbf{r}) \psi_j(0)\rangle$ decays exponentially.
More directly, we use iDMRG to calculate the energy of a charge  $e$ and charge $-e$ excitation on top of the $\nu = 2 + \delta$ ground state.
We find they are gapped throughout the SC regime: at $J = 1, \lambda = 0.5$, for example, we find a particle-hole gap of $\Delta_{\textrm{PH}} \approx 0.61 E_C$ at $\nu = 2 + 1/4$ and $\Delta_{\textrm{PH}} \approx 0.55 E_C$ at $\nu = 2 + 1/2$, independent of the layer and isospin index of the added charges (see App. \ref{appss:PHgap}).
In contrast, the charge-$2e$ excitations must be gapless by virtue of the algebraic correlations in $\Delta(\mathbf{r})$.

\begin{figure}
    \includegraphics[width=0.8\columnwidth]{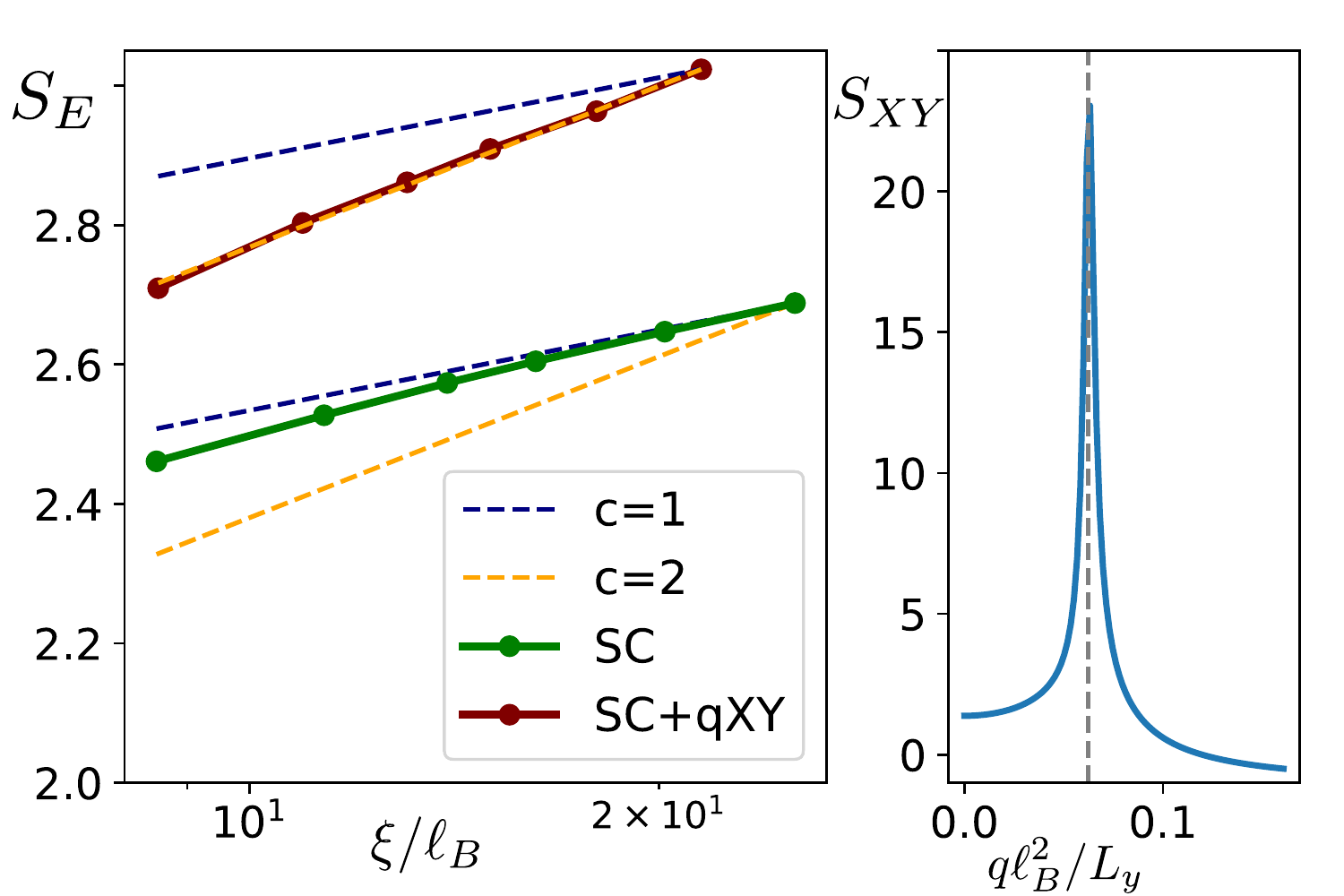}
    \caption{\textbf{a})   Determination of the central charge via finite-entanglement scaling in the superconducting ($J = 0.9, \lambda = 0.6$) and SC/qXY coexistence ($J = 0.55, \lambda = 0.2$) phases at doping $\delta = \frac{1}{4}$. By tracking the increase of the entanglement entropy $S_E$ and MPS correlation length $\xi$ with the DMRG bond dimension $\chi$, we extract the central charge from the scaling relation $S_E = \frac{c}{6} \log(\xi/a)$ \cite{PollmannFES2009}. Dashed lines show the expected slopes for $c=1, 2$. The SC/qXY phase fits $c=2$ perfectly, while the SC phase approaches $c=1$ at the largest length scales. The larger slope observed for small $\xi$ is consistent with the XY order of the SC/qXY phase being destroyed at a continuous (Kosterlitz-Thouless) transition. 
    \textbf{b}) Fourier-transform of the XY-correlation function $S_{\textrm{XY}}(q) = \langle N^{+}(q) N^{-}(0) \rangle$ in the SC/qXY coexistence phase. The XY order is shifted to finite wavevector $q_\ast = \delta L_y / 4 \ell_B^2$, where $\delta$ is the doping. This wavevector is consistent with a fluctuating meron gas. }
    \label{fig:central_c}
\end{figure}

As a final confirmation of superconductivity, we use ``finite entanglement scaling'' \cite{PollmannFES2009} to extract the central charge $c$ of the effective 1D model, Fig.~\ref{fig:central_c}.
Throughout most of the SC, we find $c=1$, consistent with fluctuations of the SC phase mode $\phi$ but no other gapless fermionic or bosonic excitations.
The only exception is in a region close to the CDW/SC transition, where we find $c=2$. 
As we will later discuss, in this region the SC coexists with a finite wavevector version of the XY-order (qXY in Fig.~\ref{fig:phase_diagram}), with fluctuations in $\phi, \theta_{\textrm{XY}}$ contributing $c = 1$ apiece.

\textit{Skyrmions.} It is already remarkable that we find superconductivity in a purely repulsive model.
But how do we tell whether this SC is related to skyrmions?
To explore this question, it will prove helpful to  review the NL$\sigma$M description of Eq.~\eqref{eq:H},  which predicts the existence of charge-$2e$ skyrmions which we can then quantitatively compare against our DMRG numerics.
Consider first a single layer (say $\gamma^z = \pm$), with $\nu_{\pm} = 1$ of its two LLs filled.
In isolation each layer is analogous to a spinful QH system at $\nu_T = 1$, with ferromagnetic order parameter $\mathbf{n}_{\pm}(r) = 2 \pi \ell_B^2 \langle \psi^\dagger_{\pm}(r) \vec{\eta} \psi_{\pm}(r) \rangle$ given by its isospin polarization.
According to the theory of quantum Hall ferromagnetism,\cite{Sondhi,MoonMori,girvinLesHouches} fluctuations in $\mathbf{n}_{\pm} \in S^2$ are governed by the Lagrangian 
\begin{align}
\mathcal{L}^{\textrm{QHFM}}_{\pm}  &= \int_r \frac{1}{2 (2 \pi \ell_B^2)} \mathbf{\mathcal{A}}_\pm \cdot \partial_\tau \n_\pm + \frac{g}{2}  (\nabla \mathbf{n}_{\pm})^2 + \mathbf{A}_{\mu} \cdot \mathbf{j}^\mu_\pm \notag \\
& \quad + \frac{1}{2}\int_{r,r^\prime} \rho_{\pm}(r) V_C(r - r^\prime) \rho_{\pm}(r^\prime), \notag \\
\mathbf{j}^\mu_{\pm} & = \pm\frac{e}{8 \pi} \epsilon^{\mu \nu \rho} \mathbf{n}_{\pm} \cdot \left( \partial_\nu \mathbf{n}_{\pm} \times \partial_\rho \mathbf{n}_{\pm} \right)
\label{eq:L_QHFM}
\end{align}
where $\mathbf{\mathcal{A}}_\pm[\n_\pm]$ corresponds to the \textit{vector potential} for the isospin-half Berry phase \cite{girvinLesHouches}, $\mathbf{A}$ is the external vector potential that couples to the electric current density $\mathbf{j}^\mu_\pm$, and the isospin stiffness is $g = \frac{\ell_B^2}{32 \pi^2} \int_0^\infty dq \, q^3 e^{-q^2/2} V_C(q)$ \cite{}.
The key feature is that textures in $\mathbf{n}_{\pm}$
induce an electric charge density through the relation
$\rho_{\pm}= C_{\pm} \frac{e}{4 \pi}   \mathbf{n}_{\pm} \cdot \partial_x \mathbf{n}_{\pm} \times \partial_y \mathbf{n}_{\pm}$, where $C_{\pm} = \pm $ is the Chern number.
The reason for this is that as an electron moves through the system, its isospin cants to follow the texture $\mathbf{n}_{\pm}$, generating a Berry phase.
The electron responds to the Berry phase just like a magnetic field, and so the resulting Berry curvature is converted into electric charge via the Hall response $\sigma_H = C \frac{e^2}{h}$.
Integrating this relation one finds $Q = C Q_{\textrm{topo}}$, where $Q_{\textrm{topo}}$ is the total skyrmion number and $Q$ is the total charge. 
The long-range part of the Coulomb repulsion $V_C$ then prefers to make large skyrmions in order to spread out the charge, lowering the skyrmion energy relative to the bare electron's.
The lowest energy charged excitations of a QHFM are thus charge-$1e$ skyrmions, which has been well established experimentally in a variety of QH systems.\cite{Tycko,Tycko2,West}
 
 When considering two layers with opposite $C$, we can  extend Eq.~\eqref{eq:L_QHFM} by  coupling the layers through the anti-ferromagnetic interaction $\bar{J^i} = J^i E_C/(2 \pi A_M)$ and Coulomb repulsion,
\begin{align}
  \mathcal{L} &= \int_r   \sum_{\gamma} \left[ \frac{1}{2 (2 \pi \ell_B^2)} \mathbf{\mathcal{A}}_\gamma \cdot \partial_\tau \n_\gamma + \frac{g}{2} (\nabla n_{\gamma})^2 + \mathbf{A}_{\mu} \cdot \mathbf{j}^\mu_\gamma \right] \notag \\
&+ \frac{1}{2} \int_{r,r^\prime} \sum_{\gamma, \gamma'} \rho_{\gamma}(r) V_C(r - r^\prime) \rho_{\gamma'}(r^\prime) - \frac{J^i}{(2 \pi \ell_B)^2} \int_r (\mathbf{n}^i_+ - \mathbf{n}^i_-)^2   
\label{eq:NLSM2}
\end{align}
The behavior of the skyrmions in this model is quite rich, depending on $g / J$ and $\lambda / J$.
Skyrmions in layer-``$+$'' carry charge $+1$, while skyrmions in layer-``$-$'' carry charge $-1$, so 
$Q = Q^+_{\textrm{topo}} - Q^-_{\textrm{topo}}$.
When an electron is added to each layer,  they thus enter as a skyrmion in layer ``$+$'' and an \emph{anti}-skyrmion in layer ``$-$''.
What is the effective interaction between them?
 $V_C(r)$, of course, would like to push the two charge-$e$ objects apart.
 However, if the skyrmions separate then there are regions in which the $\mathbf{n}_+, \mathbf{n}_-$ fields are no longer anti-ferromagnetically aligned, costing $J$.
If they instead sit right on top of each other $J$ is always satisfied because the anti-skyrmion solution is obtained by flipping the spin of a skyrmion solution, $\mathbf{n}_{\textrm{a-skyr}} = -\mathbf{n}_{\textrm{skyr}}$, thereby generating an attractive interaction.
 
 Remarkably, if $\lambda = 0$, a careful analysis of Eq.~\eqref{eq:NLSM2} in App.~\ref{App:NLSM} (see also Refs.~\onlinecite{Eslam,Chatterjee19}) shows that charge-1e skyrmions will prefer to bind into a single charge-2e skyrmion for \emph{any} $J > 0$. This is because the skyrmion can spread out over an arbitrarily large radius $R$. Since the Coulomb repulsion falls off as $V_C \sim \frac{1}{R}$, $J$ eventually wins out. 
 The situation is more complicated when $\lambda > 0$, where the easy-plane anisotropy deforms the skyrmion into a meron pair.
 Roughly speaking, this contributes an elastic energy $g \log(R)$ to the object, cutting off its maximal size $R$. 
 In this case, there is a finite critical $J \geq J_\ast(\lambda)$ where the attraction wins out.
 
\begin{figure}
    \includegraphics[width=\columnwidth]{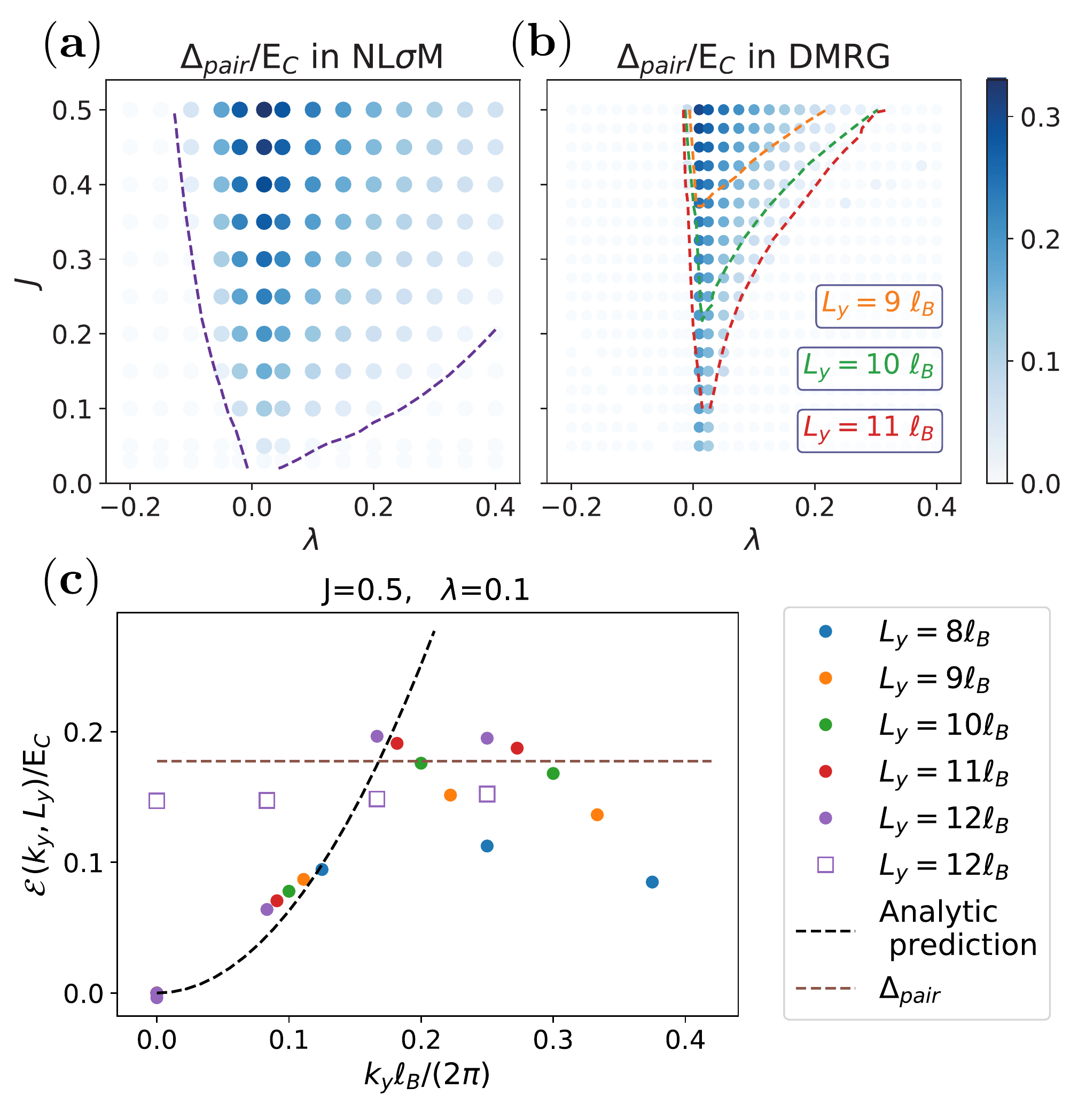}
    \caption{
    $\Delta_{\textrm{pair}}(J,\lambda)$ evaluated numerically using the classical NL$\sigma$M (\textbf{a}) and quantum DMRG (\textbf{b}) show qualitative agreement. In (\textbf{a}), the $2e$ bound state is preferred in the blue region ($\Delta_{\textrm{pair}} > 0$), demarcated by dotted purple lines ($L_x,L_y = 21 \ell_B$). In (\textbf{b}), the blue background indicates $\Delta_{\textrm{pair}}$ extrapolated to the $L_y \to \infty$ limit, while the dashed lines show non-extrapolated contours of $\Delta_{\textrm{pair}} = 0$ for different $L_y$. (\textbf{c}) Solid-circles denote $\mathcal{E}_{2e}(k_y, L_y) \equiv E_{2e}(k_y,L_y) - E_{2e}(0,L_y)$ for a typical $(J, \lambda)$ in the superconducting phase, showing that charge-$2e$ excitations disperse at small $k_y$ with an effective mass that agrees reasonably well with the classical estimate (dashed black line). In contrast, $\mathcal{E}_{1e}(k_y, L_y) \equiv \sum_{\gamma = \pm} E_{1e,\gamma}(k_y, L_y) -  E_{2e}(0, L_y)$ (purple squares) shows that the charge-$1e$ dispersion is flat. For large $k_y, L_y$, both approach the $L_y$-extrapolated value of $\Delta_{\textrm{pair}}$ (dashed brown line), indicating that for $k_y \ell_B / (2 \pi) \gtrsim 0.2$, the $2e$-pair disassociates into two well-separated $1e$ excitations.
    }
    \label{fig:Excitations}
\end{figure}

To quantitatively understand the energetics of the skyrmion binding in the $(J, \lambda)$ plane, we  solve for the lowest-energy charged excitations of Eq.~\eqref{eq:H} at $\nu=2$ both semiclassically and using DMRG. 
First treating Eq.~\eqref{eq:NLSM2} classically, we  numerically solve for its ground state in order to compute the pair binding energy $\Delta_{\textrm{pair}} = 2 E_{1e} -  E_{2e} $, where $E_{2e}$ is the energy of the charge $2e$ skyrmion / anti-skyrmion pair, while $2 E_{1e}$ is the energy of a well-separated $1e - 1e$ pair.  
In Fig.~\ref{fig:Excitations}a, we see that $\Delta_{\textrm{pair}}(J, \lambda)$ has a fan-like structure within which the interaction is attractive,  reminiscent of the region where we observe superconductivity upon doping. In particular, letting $J_\ast(\lambda)$ denote the critical value of $J$ required for pair formation at a given anisotropy $\lambda$, we see that $J_\ast(\lambda \to 0) \to 0$.
Pair formation is more favorable on the easy-plane side, requiring a smaller $J_\ast$ for the same absolute value of $\lambda$. The physical reason is that for $\lambda > 0$ the 2e pair can deform into a topologically equivalent texture -- a confined pair of charge-$e$ merons \cite{ActorRMP,MoonMori} with well-separated cores -- thereby lowering the electrostatic charging energy at the expense of an additional elastic energy cost  which is quantitatively small (note $g \approx 0.025 E_C$ ).
This deformation mechanism is not allowed for easy-axis 2e skyrmions, resulting in a steeper slope for $J_\ast$ in the $(J,\lambda)$ plane.

We next go beyond the NL$\sigma$M by computing the  energies of skyrmion excitations using DMRG.
To do so, we start with the DMRG ground state of Eq.~\eqref{eq:H} at $\nu = 2$.
We then consider  an excitation with either a single charge in one layer (1e), or two charges, one in each layer (2e). 
Because they are distinguished by their quantum numbers from each other and the $\nu= 2$ vacuum, DMRG can be used to target the lowest energy state in each quantum number sector, resulting in DMRG energies $E_{1e}(k_y), E_{2e}(k_y)$ measured relative to the vacuum, where $k_y$ is the momentum around the cylinder. 
The excitation energies are obtained from infinite-cylinder DMRG using the approach of Ref.~\onlinecite{zaletelpollmannmongQH2013}. For numerical details including the convergence with $L_y$ and $\chi$ we refer to App.~\ref{App:SegDMRG}.
A typical ``dispersion relation'' for a $J, \lambda$ exhibiting superconductivity is shown in Fig.~\ref{fig:Excitations}c.
We see that $E_{1e}(k_y)$ is exactly flat, reflecting that this excitation sees a net magnetic field which quenches its motion.
$E_{2e}(k_y)$, in contrast, is dispersive and shows a minimum at $k_y=0$.
Our $k_y$-resolution is too coarse to extract a dispersion relation,\footnote{While it is tempting to extract a dispersion relation by interpolating between different $L_y$, we see from the $k \gg \ell_B^{-1}$ limit that the energies show a residual $L_y$ dependence at fixed $k$, which is expected because of the $1/r$ Coulomb interaction. So it is best to leave this analysis at the level of an order-of-magnitude comparison.}
but for comparison we plot the expected dispersion of a $2e$-skyrmion pair using a classical estimate \cite{Eslam} for the effective mass $m$,  $\left(\frac{\hbar^2 \ell^{-2}_B}{2m}\right) /E_C =  J/\pi$, and find that for our largest $L_y$ they agree to within 20\% for small anisotropy $\lambda \lesssim 0.1$.
For $k_y \ell_B \gtrsim 1.5 $ the energy saturates at $E_{2e}(k_y) \to 2 E_{1e}$, indicating that the pair disassociates.
We caution the reader that we only expect order of magnitude agreement between the classical and quantum results because the mass will be corrected by quantum fluctuations, as analyzed in App.~\ref{appss:compare}. 

\begin{figure*}
     \centering
     \begin{subfigure}[b]{0.49\textwidth}
         \centering
         \includegraphics[width=\textwidth]{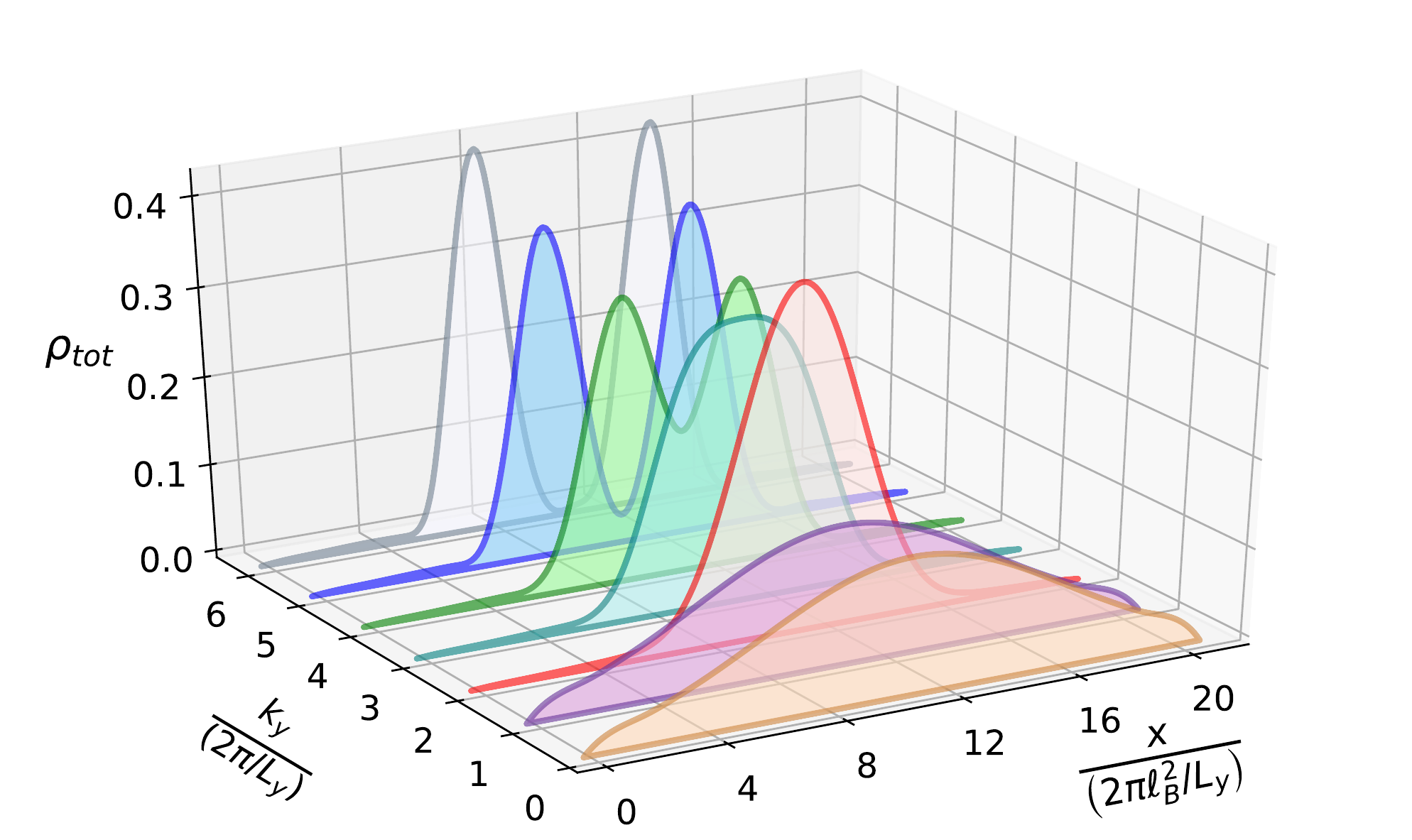}
         \caption{Total charge density $\rho_{tot} = \sum_{\gamma = \pm} \rho_\gamma$}
         \label{fig:NetCD}
     \end{subfigure}  
     \hfill     
     \begin{subfigure}[b]{0.49\textwidth}
         \centering
         \includegraphics[width=\textwidth]{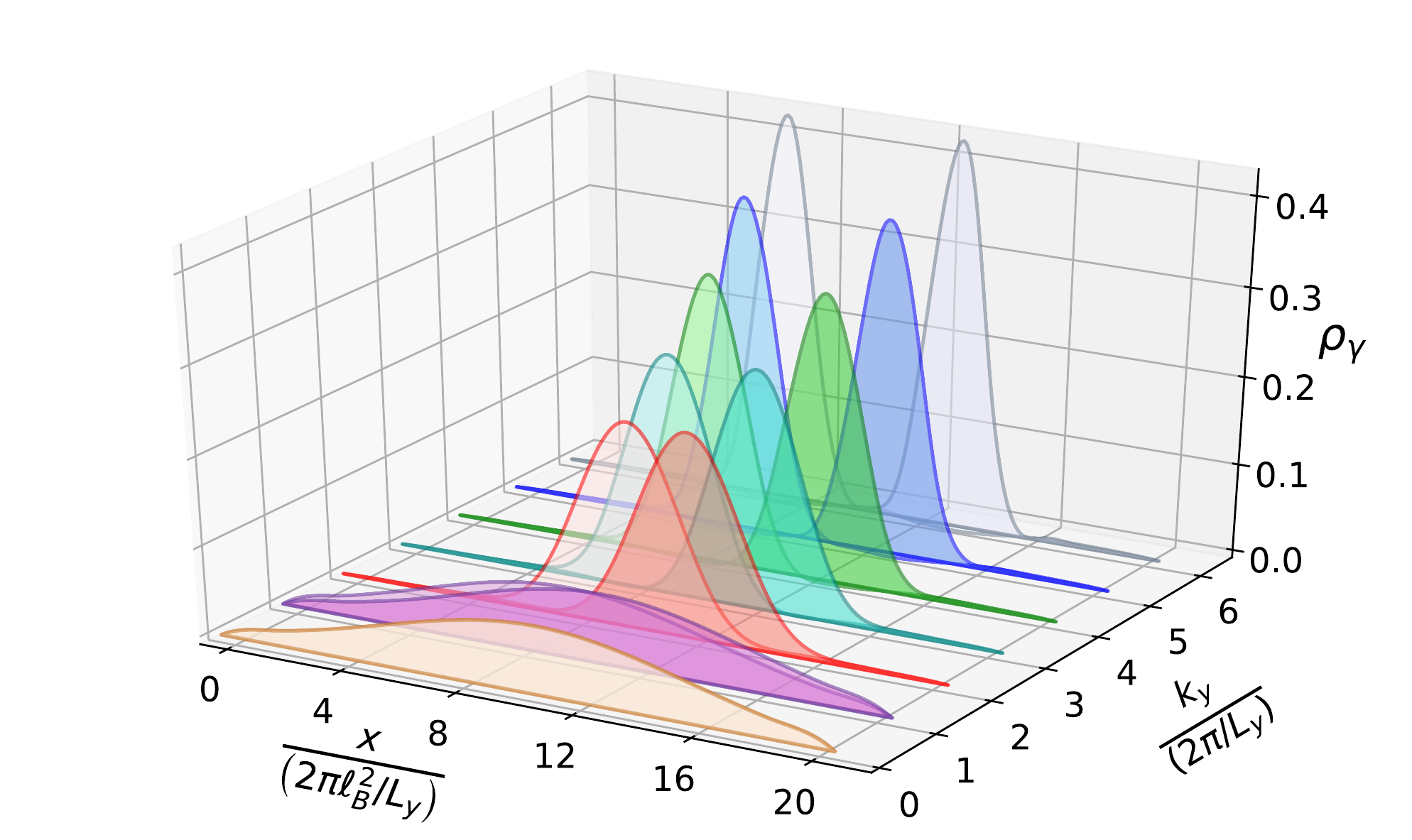}
         \caption{Layer resolved charge densities $\rho_\gamma$ (light/dark hues for $\gamma = \pm$)}
         \label{fig:dipoleCD}
     \end{subfigure}
        \caption{Evolution of charge densities (net and layer-resolved) of a charge $2e$ excitation as a function of momenta $k_y = 2 \pi k/L_y$ at $(J,\lambda) = (0.5,0.1)$ and $L_y = 12 \ell_B$. Note that at $k_y = 0$, the charges in the two layers are exactly on top of each other, but at finite $k_y$ they move away with a separation $\Delta x = k_y \ell_B^2$, thereby losing exchange energy.}
        \label{fig:dipole}
\end{figure*}

We see both the ingredients for superconductivity:  first pairing ($E_{2e}(0) < 2 E_{1e}$), and second, despite the completely quenched band dispersion of the electrons, the pairs have a disperse with scale $J$ so can support a finite superfluid stiffness.
There is a beautiful explanation for the finite dispersion which gives an intuitive picture for the previously-discussed  lower bound on the superfluid stiffness.\cite{Andrei, Peotta_2015, Hazra}
In close analogy to Gorkov and Dzyaloshinskii's analysis of a Mott exciton in a finite-field, \cite{gor1968contribution} because the two charges in the pair have opposite Chern number (e.g. $B$-field), when the pair drifts at velocity $\mathbf{v}$ the charges feel equal and opposite Lorentz forces $\pm e |B|  \mathbf{v} \times \hat{z}$, pulling them apart.
This force is counteracted by the pairing attraction $\Delta_{\textrm{pair}}(r)$, where $r$ is the distance between the pair.  Equating $|e B v| =  \partial_{r} \Delta_{\textrm{pair}}(r)$, and defining $m$ as $\Delta_{\textrm{pair}}(r) \approx \Delta_{\textrm{pair}}(0) + \frac{\hbar^2 r^2}{2 m \ell_B^{4}}$, 
we find $\mathcal{E}_{\textrm{pair}}(v) = \Delta_{\textrm{pair}}(0) + \frac{1}{2} m v^2$.
In a Bloch band, an equivalent result can be obtained from the  $k$-space Berry curvature $\Omega(\mathbf{k})$ using the semiclassical relation $\mathbf{v} = -\frac{1}{\hbar} \nabla_{\mathbf{r}} \Delta_{\textrm{pair}}(\mathbf{r}) \times \Omega(\mathbf{k})$.
Either way, $m$ (and hence $\rho_{\textrm{SC}}$) is generated entirely by the interplay of the interaction $\Delta_{\textrm{pair}}$ and the Chern number,  and hence the two ingredients for superconductivity always come in tandem.
Since $\mathbf{r} = \ell_B^2 \hat{z} \times \mathbf{k}$ by these relations, the large $k$ limit rips apart the pair, explaining the limit  $E_{2e}(k_y) \to 2 E_{1e}$.
Note that this same mechanism is familiar in more conventional quantum Hall contexts: it gives rise to the $E_C$-scale mass of composite fermions at $\nu = \frac{1}{2}$\cite{pasquier1998dipole} and excitons in quantum Hall bilayers \cite{YangDipolar2001}. 
This intuitive picture is confirmed by our numerical computation of layer-resolved charge densities $\rho_{\gamma}$ ($\gamma = \pm$) for a $2e$ excitation as a function of $k_y$, as shown in Fig.~\ref{fig:dipole}.
On inserting charge $2e$ at momentum $k_y = 0$ on top of the insulating state, the additional charge density in each layer lies exactly on top of each other, forming a charge $2e$ bound state --- the $2e$ skyrmion.
As $k_y$ is increased, this $2e$ excitation unbinds into two charge $e$ excitations.

We next use DMRG to compute $\Delta_{\textrm{pair}}$ at $k_y = 0$ as a function of parameters $J, \lambda, L_y, \chi$.
We extrapolate the energies with respect to $L_y$ and $\chi$ to obtain the pair binding energy $\Delta_{\textrm{pair}}(J, \lambda)$ (see App.~\ref{App:SegDMRG}). 
The result, shown in Fig.~\ref{fig:Excitations}b, is in good qualitative agreement with the NL$\sigma$M results.  
Specifically, $J_\ast(\lambda)$ vanishes as $\lambda \to 0$, and pair-formation requires smaller  $J$ in the easy-plane case.
One quantitative discrepancy is  the  $J_\ast(\lambda)$ boundary found in DMRG is shifted to higher $J$ relative to the NL$\sigma$M.
In App.~\ref{appss:compare}, we show that this effect can be qualitatively reproduced by adding quantum zero-point fluctuations to the NL$\sigma$M, which increase the energy of the charge $2e$ excitations by an amount proportional to $J$ in the small anisotropy limit,  which deters pairing.

The behavior of $\Delta_{\textrm{pair}}(J, \lambda)$ supports the following explanation for the phase diagram  at $\nu = 2 + \delta$ for low-$\delta$, Fig.~\ref{fig:phase_diagram}.
The layer-polarized CDW region corresponds to the case where $\Delta_{\textrm{pair}} < 0$: charges enter as well-separated electrons and form a CDW pattern for the same reason low-density QH systems are known to form various Wigner-crystal and stripe phases.\cite{Fogler} 
The SC region corresponds to the regime where $\Delta_{\textrm{pair}} > 0$: charges instead enter as charge-$2e$ meron-pairs and condense.

An apparent discrepancy in this interpretation  we should first address is  the critical $J_c$  of the doped CDW/SC transition at $\lambda = 0$.
For the doped phase diagram  $J_c(\lambda=0, \delta = 1/4, L_y = 10 \ell_B) \sim 0.3$, while for $\Delta_{\textrm{pair}}$ we found that $J_\ast(\lambda = 0) = 0$.
This is actually an expected finite size effect. Note $\Delta_{\textrm{pair}}$ was extrapolated to $L_y \to \infty$, while the doped phase diagram is shown at fixed $L_y = 10 \ell_B$ because it was difficult to doubly extrapolate $\chi, L_y \to \infty$ from the available data (see App.~\ref{App:LyJc}). 
So in Fig.~\ref{fig:phase_diagram} and \ref{fig:Excitations}b we also demarcate  the $J_\ast(\lambda, L_y = 10 \ell_B)$ boundary without  $L_y$ extrapolation, and indeed we find $J_\ast(\lambda = 0, L_y = 10 \ell_B) \approx 0.25$, in decent agreement with $J_c$.
The origin of the finite-size effect is the behaviour of the NL$\sigma$M on a cylinder. 
In 2D, the elastic energy of a skyrmion is scale invariant, so at $\lambda = 0$ it can grow to arbitrary size in order to reduce its Coulomb energy. 
On a cylinder, however, an analytic solution of the NL$\sigma$M shows that the skyrmion cannot grow beyond $R \sim L_y$, lower-bounding  it's Coulomb energy by $\sim V_C(L_y)$. 
Thus at finite $L_y$, a finite $J$ is required to overcome Coulomb repulsion.
However note that as $\delta$ increases, we don't expect such quantitative agreement between $J_\ast, J_c$, because the inter-skyrmion interaction energy becomes important in addition to $\Delta_{\textrm{pair}}$; this discrepancy is seen for the large-$\delta$ phase diagram, App.~\ref{App:doping}.

Further evidence for skyrmion pairing can be gleaned from the region where the SC order coexists with a finite-wavevector XY-order, with 1D central charge $c=2$ (Fig.~\ref{fig:central_c}a). 
We call this the qXY-order because the wavevector of the XY-order is shifted to a finite $q_\ast$ along the cylinder, with leading behavior
\begin{align}
 \langle N^{+}(x, 0) N^{-}(0, 0) \rangle \sim x^{-\eta_{\textrm{XY}}} \cos(q_\ast x) + \cdots \label{eq:qXY}
\end{align}
for $x \gg L_y$, as  shown in Fig.~\ref{fig:central_c}b. 
Within the range of $L_y, \delta$ we have explored, we find $q_\ast$ is always locked to the doping $\delta$ according to the relation $\frac{\delta}{2 \pi \ell^2_B} L_y \frac{2 \pi}{q_\ast} = 4$.
In other words, $\theta_{\textrm{XY}}(x)$ increments by $\Delta \theta_{\textrm{XY}} = \pi$ every time $x$ passes charge-$2e$ worth of doping.

This curious effect is in fact further evidence for skyrmion superconductivity. 
Recall that in the easy-plane regime, the $2e$-skyrmion deforms into a pair of bound merons, each carrying charge $1e$.
In terms of $\theta_{\textrm{XY}}$, this object is a vortex / anti-vortex pair.
In the easy-plane limit, the elastic energy is $E = \frac{g}{2} \int d^2r (\nabla \theta_{\textrm{XY}})^2$, and we can solve for the field configuration $\theta$ which minimizes $E$ subject to the constraint of unit vorticity at $z_0 = x + i y$ and anti-vorticity at $z_1$. Using a conformal transformation to map the solution of the Laplace equation from the plane to the cylinder, we find that
\begin{align}
\theta(z)  &= \arg\left [ \sinh(2 \pi (z - z_0) / 2 L_y )  \sinh(2 \pi (\bar{z} - \bar{z}_0) / 2 L_y ) \right ] \notag \\
\Delta \theta &= \theta(x = \infty, y) - \theta(-\infty, y) = 2 \pi (y_1 - y_0) / L_y.
\end{align}
We see that the phase jumps by an amount $\Delta \theta$ in proportion to vertical  displacement $\Delta y = y_1 - y_0$ between the merons. 
Because of the Coulomb repulsion, for small $\lambda$ the meron-pair will prefer to spread across  the circumference of the cylinder, $\Delta y = L_y/2$, corresponding to $\Delta \theta = \pi$.
So \emph{if} the doping $\delta$ enters as large meron-pairs, $\theta_\textrm{XY}(x)$ should jump by $\pi$ per  $2e$ passed along the cylinder. 
This is exactly the wavevector $q_\ast$ we observe in the SC region.
In contrast, in the CDW phase, the XY wavevector remains at $q = 0$, consistent with the charge $\delta$ entering as electrons. 

One might object that if the SC is a condensate of meron-pairs, then the resulting $\pi$-fluctuations in $\theta_{\textrm{XY}}$ would immediately destroy the XY-order.
But this is not the case on a cylinder geometry because the SC order is  algebraic, as we demonstrate using bosonization in App.~\ref{app:qXYbosonization}.
However, as the SC stiffness increases the density fluctuations become larger and the XY order is eventually destroyed at a BKT transition.
This is consistent with the absence of qXY-order for large-$J$ (Fig.~\ref{fig:phase_diagram}), where the central charge flows from $c=2 \to 1$ (Fig.~\ref{fig:central_c}a). 
This also explains how $q_\ast$ can depend on $L_y$, which would otherwise appear to be unphysical in the 2D limit: the SC stiffness increases linearly with circumference, so as $L_y \to \infty$ the width of the qXY order shrinks to zero.
So the qXY does not exist as a 2D phase, but rather as a unique fingerprint of the skyrmion SC when placed on the cylinder geometry. 

\textit{A control experiment.} Finally, we confirm the role of topology using  a ``control'' experiment: we consider a Hamiltonian identical to Eq.~\eqref{eq:H}, but with all four components $\psi_{\gamma \eta}$ in the same magnetic field.
The ground state at $\nu = 2$ is still found to have XY-order, so the system still admits skyrmions in each layer independently. 
However the skyrmion-pairing mechanism we have identified is inoperative because a charge-2e excitation now requires the \emph{same} skyrmion handedness in each layer, so $J$ does not generate attraction. 
Running iDMRG for the same $J = 0.9, \lambda = 0.6$ where the opposite-$B$ model is a strong superconductor, we find the SC correlations now decay exponentially (by three orders of magnitude per $\ell_B$).\footnote{In the same-$B$ case superconductivity requires an Abrikosov vortex lattice, which enlarges the unit cell. We account for this possibility by running iDMRG with the requisite unit cell both around and along the cylinder so that we don't spuriously forbid a SC.} 
Note that in the control scenario  \textit{ferromagnetic}  exchange ($J<0$) could favor the formation of charge-2e skyrmion-pairs, which has in fact been argued to occur in certain conventional quantum Hall systems.  \cite{Lilliehook,Abanin,NK_PRL98}. 
However, such a pair experiences a net magnetic field,  leading to a flat dispersion which makes superconductivity via condensation unlikely. 

In conclusion, we have shown that a model capturing the symmetry and topology of twisted bilayer graphene features a novel all-electronic route to superconductivity.
The ``mother state'' of the superconductor is an XY-order whose lowest-energy charged excitations are charge-2e skyrmions, despite the long-range Coulomb interaction.
When doped, the finite  density of skyrmions Bose condense and form a superconductor. 

It is worth commenting on the relation of our findings to the proposal of Grover and Senthil,\cite{GroverSenthil} which was recently explored numerically, for example, in Ref.~\onlinecite{Assaad}, where it was found that doping an interaction-driven quantum spin Hall (QSH) state lead to a SC.
From a topological point of view, this mechanism is analogous to the one discussed here  under the identification of our XY-order with the QSH state (and setting our $\lambda = 0$). 
However, energetically, the model which was studied keeps the analog of our ``$J$'' term (which generates the QSH state), but does not contain the Coulomb repulsion $V_C$, which, unfortunately, would lead to a sign problem for determinantal quantum Monte Carlo. On its own, $J$ can be decoupled into an attractive interaction which then has no competition with $V_C$, so superconductivity is stabilized at the mean-field level.  
This is not to say our work disagrees with their conclusions, but, by explicitly showing that the skyrmion energetics at integer filling is predictive of superconductivity upon doping, we  demonstrate that skyrmion pairing is necessary and sufficient for superconductivity and is robust to both $V_C$ and $\lambda > 0$.

   Where does MATBG lie in the phase diagram? We can very roughly estimate the values of $J, \lambda$ realized in MATBG using the relation $A_M = 2 \pi \ell_B^2$. From the MATBG Hartree-Fock results of Ref.~\onlinecite{NickPRX}, which computed the energy of the layer ferromagnet ( ``VP'') phase relative to the layer anti-ferromagnet (``KIVC'' and ``VH'') phases, we then find $\lambda \sim 0.1$ and $J \sim 0.05 - 0.3$ for a dielectric constant of $\epsilon = 10 \epsilon_0$, depending on details like the twist angle and gate distance. 
    It is thus quite feasible that MATBG is in the regime where the lowest energy excitations are charge-$2e$ skyrmions.
However, there are some important quantitative differences between MATBG and the model studied here. These include the narrow band dispersion (though its most significant effect is already included via the generation of the superexchange $J$ between layers\cite{NickPRX}), and the inhomogeneity of the Berry curvature, so this comparison should be made with caution.
   Future DMRG studies of the MATBG Hamiltonian could help decide the issue.  \cite{KangVafek, SoejimaParker}

    More broadly, while  our model is inspired by the physics of MATBG,
    the basic ingredients of skyrmion superconductivity are simple: two spinful (or isospinful) bands with opposite Chern number. Might these ingredients already be out there in other solid state systems?
    Alternating angle twisted trilayer graphene, which has identical low-energy topological bands as MATBG \cite{Khalaf2} and has recently been shown to display robust superconductivity  \cite{Park2021,Hao2021}, offers another possible material candidate for such a skyrmionic mechanism.
    Furthermore, a system where $E_C$ was at the atomic, rather than moir\'e scale, would provide a new route to high-temperature superconductivity.

\acknowledgements{
We acknowledge our collaboration with Nick Bultinck, Eslam Khalaf, and Ashvin Vishwanath which was the inspiration for this work.
M.Z. thanks  Norman Yao and Andrea Young for helpful conversations, and Roger Mong and Frank Pollmann for codevelopment of the TenPy DMRG code. 
S.C. was supported by the ARO through the Anyon Bridge MURI program (grant number W911NF-17-1-0323), and the W. M. Keck Foundation via Norman Y. Yao.
M.I. was supported by the Gordon and Betty Moore Foundation’s EPiQS Initiative through Grant GBMF4302 and GBMF8686.
M.Z. was supported by the Office of Basic Energy Sciences, MSE Division of the U.S. DOE under contract no. DE-AC02-05-CH11231 (van der Waals heterostructures program, KCWF16). 
Computations were performed on Stanford Research Computing Center's Sherlock cluster and the Lawrencium computational cluster resource provided by the IT Division at the Lawrence Berkeley National Laboratory (Supported by the Director, Office of Science, Office of Basic Energy Sciences, of the U.S. Department of Energy under Contract No. DE-AC02-05CH11231).}

\bibliography{refs}

\appendix
\onecolumngrid

\section{Methods}
\label{App:iDMRG}

Here we detail the application of infinite DMRG to the Hamiltonian Eq.~\eqref{eq:H} and  the observables used to determine the phase diagram.
In order to apply our existing DMRG algorithms, it is technically convenient to first apply a unitary  PH-symmetry transformation to the two $C=-1$ bands of Eq.~\eqref{eq:H}, mapping $\psi_{-, i} \to i \eta^y_{ij} \psi^\dagger_{-, j}$. Because $C$ is odd under a unitary PH transformation, this  maps the problem onto a conventional quantum Hall bilayer (i.e., one where both layers have Chern number $C=+1$). 
This transformation yields an exact rewriting of Eq.~\eqref{eq:H} as
\begin{align}
    H 
    & = \psi^\dagger \frac{(\mathbf{p} + e \mathbf{A})^2}{2m} \psi + 
    \frac{1}{2} \int : \psi^\dagger(r) \gamma^z \psi(r) V_C(r-r') \psi^\dagger(r') \gamma^z \psi(r') :  \\
    & - E_C \ell_B^2 \sum_{i=x,y,z} J_i : (\psi^\dagger(r) \gamma^z \eta^i \psi(r))^2 : + \alpha \hat{N} + \beta \hat{P} + \gamma N_\phi 
    \label{eq:PH_hamiltonian}
\end{align}
Here $\hat{N}$ is the total charge, $\hat{P}$ is the total layer polarization, and $N_\phi$ is the number of flux quanta. The single-particle shifts $\alpha, \beta, \gamma$ arise from the commutators required to bring $H$ back to  normal-ordered form after the PH transformation, and they can be computed analytically from $V_C, J_i$. 
Notice that the kinetic term is now $\gamma^z$ independent,  but Coulomb energy  depends on $\psi^\dagger \gamma^z \psi$, i.e., the difference between densities in the two layers, rather than the conventional density $\psi^\dagger \psi$.

Eq.~\eqref{eq:PH_hamiltonian} then represents a traditional multicomponent quantum Hall problem, albeit with a peculiar form of Coulomb repulsion.
The problem can thus be projected to the zeroth Landau level (ZLL) assuming sufficiently large energy gaps to the higher Landau levels. 
After ZLL projection, the kinetic term is quenched, the contact interactions $J_i$ are implemented as Haldane $V_0$ pseudopotentials with appropriate component indices,
and the Coulomb interaction $V_C$ is modified by the ZLL form factor.
In this form, the problem can be tackled with iDMRG~\cite{zaletelmulticomponent} by placing the system on an infinitely long cylinder of circumference $L_y$.

The iDMRG method has two built-in cutoffs: the finite  cylinder circumference $L_y$, and the size  ``$\chi$'' of the matrix product state used to approximate the ground state. The bipartite  entanglement of the MPS ansatz is bounded by $S_E  \leq \ln\chi$, while gapped ground states have area-law entanglement entropy ($S_E \propto L_y$ in our case), $\chi$ should increase exponentially in $L_y$ to maintain a desired level of accuracy.
This is the main numerical limitation on this approach, and is the reason why we consider $L_y\leq 12\ell_B$ in this work.

Another limitation is associated to the choice of bulk doping $\delta$. 
DMRG exactly preserves the three U(1) quantum numbers associated to charge (C), spin (S), and layer (L).
Consequently the state has three well-defined  ``filling fractions'' $\nu_{C/S/L}$ describing their quantum numbers per unit length.
For rational fillings $\nu = p/q$, the length of the unit cell of an infinite MPS is lower-bounded by the least-common-multiple of the denominators $q_{C/S/L}$. 
So, for instance, a state with equal layer doping of $\delta_{+/-} = \frac{1}{8}$
will require an MPS unit cell of at least 8. Since the time and memory requirements scale linearly with the length of the unit cell, this restricts the granularity of the $\delta$ we can feasibly explore.

We now detail the observables shown in Fig.~\ref{fig:phase_diagram}. For both the ``SC'' and ``XY'' color plots, we compute $ S_{XY/SC}(\mathbf{q}=0) = \int d^2 r \, O^\dagger(r) O(0) $, where $O = N^+ / \Delta$ respectively, at MPS bond dimension $\chi = 6000$. These quantities are not true order parameters (they are always non-zero), but quantitatively they are many orders of magnitude larger in the SC and XY phases, so are convenient heuristics for demarcating the phase boundary.  The rigorous criterion for SC or XY order is the finite-size scaling analysis of the algebraic correlations shown in Fig.~\ref{fig:correlations}, which, at the resolution of our $J, \lambda$ grid, we find perfectly correlates with the obvious jump in $S_{XY/SC}(0)$.

For the CDW order, we compute the Fourier components of the charge density along the cylinder, $n(q_x) = \int dx dy \, e^{-i q_x x} \langle n(x, y) \rangle$, and plot the magnitude of the largest $q_x \neq 0$ component.

For the qXY order, we show the finite-$q$ structure factor $S_{XY}(q_x = q_\ast, q_y = 0)$ where $q_\ast = \delta L_y / 4$. Again, this is not a true order parameter, but the quantitative jump in this quantity correlates with a scaling analysis of the singularity in $S_{XY}(q)$ at $q_\ast$ which can be seen in Fig.~\ref{fig:central_c}b.

Finally, a fifth quantity (not shown) is the layer polarization $\nu_+ - \nu_- = 0$ or $\delta$. The polarized case perfectly correlates with XY / CDW phase, while the SC is unpolarized.

\section{Repulsive nature of the bare interaction \label{app:bare_repulsion}}
Due to the anti-ferromagnetic interaction, electrons in components $\psi_{+, \uparrow}, \psi_{-, \downarrow}$ experience an attractive $\delta(r)$-interaction from the XXZ-interaction:
\begin{align}
V_{\uparrow \downarrow}(r) = V_C(r) - 2 J E_C \ell_B^2 \delta^{(2)}(r)  
\end{align}
So if $J$ is sufficiently large, the bare interaction is attractive in the $s$-wave channel and the superconductivity would be rather trivial. 
Here, we show that  the range of $J$ considered in this work is far below this critical value ($J_c \sim 3.25$ for gate distance $d = 3 \ell_B$) .

To do so, we consider the problem of two electrons  with opposite magnetic field $A = \pm B (0, x)$ interacting through a central potential $V(q)$.
Note that if we apply a particle-hole transformation to \emph{one} of the particles, the problem maps onto an exciton in a uniform $B$-field field (with the sign of $V$ reversed). 
This problem was solved long ago, \cite{gor1968contribution} with the LL-projected result given in for example Ref.~\onlinecite{YangDipolar2001}, which is equivalent to our Eq.~\eqref{eq:LL_pair_dispersion}.
We repeat the equivalent derivation here without applying the PH transformation.

Projecting into their lowest LLs, where states are labelled by their Landau-gauge momentum $\mathbf{p}_y = \hbar k$, the Hamiltonian on a torus of volume $\mathcal{V}$ takes the form
\begin{align}
\hat{H} = \frac{1}{\mathcal{V}} \sum_{k_1, k_2, q}  |F(q)|^2 e^{i \ell_B^2 q_x (k_1 + k_2) } V(q) c^\dagger_{k_1 + q_y/2} c_{k_1 - q_y/2} d^\dagger_{k_2 - q_y/2} d_{k_2 + q_y/2}
\end{align}
Here $c$ is the field operator for electrons in the $+B$ field,  $d$ the field operator for  electrons in the $-B$ field, and $F(q) = e^{- \frac{1}{4} q^2 \ell_B^2}$ is the ``form factor'' of the $N=0$ LL.
In order to diagonalize the Hamiltonian, we consider the two-particle ansatz
\begin{align}
\ket{k_x, k_y} &= \frac{1}{\sqrt{N_\phi}}\sum_k  e^{i \ell_B^2 k k_x}  c^\dagger_{k_y/2 + k}  d^\dagger_{k_y/2 - k }  \ket{0} 
\end{align}
which carries momentum $k_x, k_y$.
Note that while $c, d$ separately transform under a magnetic algebra, so only their $k_y$ momentum is a good quantum number (in Landau-gauge), the composite object $c d$ sees no net field, so can be ascribed definite momentum $k_x, k_y$.
By simple state counting, the $\ket{k_x, k_y}$ are in one-to-one correspondence with $N_\phi^2$ 2-particle states of a torus. 
Hence they are eigenstates, with energy
\begin{align}
H \ket{k_x, k_y} &= \frac{1}{2 \mathcal{V} \sqrt{N_\phi}} \sum_{k, q}  |F(q)|^2  e^{i \ell_B^2 k k_x }  e^{i \ell_B^2 q_x k_y } V(q) c^\dagger_{k_y/2 + k + q_y}  d^\dagger_{k_y/2 - k - q_y } \ket{0} \\
&= \frac{1}{\mathcal{V} \sqrt{N_\phi}} \sum_{k, q}  |F(q)|^2  e^{i \ell_B^2 (k - q_y) k_x }  e^{i \ell_B^2 q_x k_y } V(q) c^\dagger_{k_y/2 + k}  d^\dagger_{k_y/2 - k}  \ket{0} \\
&= \left(\frac{1}{\mathcal{V}} \sum_{q} |F(q)|^2 V(q) e^{i \ell_B^2 (q_x k_y -  q_y k_x ) }   \right)   \ket{k_x, k_y}\\
E( \mathbf{k}) &= \mathcal{F}^{-1}[ V |F|^2  ](\ell^2_B \hat{z} \times \mathbf{k})
\end{align}
Here, we note that the exchange $k_x \leftrightarrow -k_y$ is the rotation $\hat{z} \times \mathbf{k}$, 
so we see that the dispersion is rotation of the inverse 2D-Fourier transformation $\mathcal{F}^{-1}$ of the effective potential $V(q) |F(q)|^2$.

If the potential is rotationally symmetric, 
\begin{align}
E(k) &= \frac{1}{2 \pi}\int_0^{\infty} dq q |F(q)|^2 V(q) J_0(k q \ell_B^2)
\label{eq:LL_pair_dispersion}
\end{align}
Note that if we were to drop the $|F|^2$, we would  get back the real-space potential: $E(\mathbf{k}) = V(\ell_B^2 \hat{z} \times \mathbf{k})$.
The $|F|^2$ factor just convolves this with the real-space shape of the LL wavefunction, smoothing it out over scale $\ell_B$.
The reason for this form can be understood from the guiding-center dynamics in the presence of the opposing magnetic fields. 
If both electrons are moving in parallel with velocity $\mathbf{v}$ and displacement $\mathbf{r}$, then the force $\mathbf{F}(\mathbf{r})$ due to $V$ must cancel the Lorentz force,
$\mathbf{F}(\mathbf{r}) = -e B \hat{z} \times \mathbf{v}$.
Since $\mathbf{v} = \nabla_{\mathbf{k}} E(\mathbf{k})$ while $\mathbf{F} = -  \nabla_{\mathbf{r}} V(\mathbf{r})$, we have
\begin{align}
\nabla_{\mathbf{r}} V(\mathbf{r}) = \ell_B^2 \hat{z} \times \nabla_{\mathbf{k}} E(\mathbf{k})
\end{align}
In a LL the kinetic energy is quenched and $V = E$. This is solved by fixing $\mathbf{r} = \ell_B^2 \hat{z} \times \mathbf{k}$ with $E(\mathbf{k}) = V(\ell_B^2 \hat{z} \times \mathbf{k})$.

For a $\delta$-function interaction $\ell_B^2 \delta^{(2)}(r)$, with Fourier transform $V(q) = 1$, and $\frac{1}{r}$ interaction, with $V(q) = \frac{2\pi}{q}$, the  integral can be done analytically to obtain
\begin{align}
E_\delta(k) &= \frac{e^{-k^2/2}}{2 \pi}\\
E_{\frac{1}{r}}(k) &= \sqrt{\frac{\pi}{2}} e^{-k^2 / 4} I_0(k^2/4)
\end{align}
In this case, we find that $E_{\frac{1}{r}}(k) - 2 J  E_\delta(k) > 0$ for all $k$ so long as $\sqrt{\frac{\pi}{2}} - \frac{J}{\pi} > 0$, giving $J_c = \sqrt{\pi^3 / 2} \sim 3.9$.
For a gate screened interaction, $V(q) = \frac{2 \pi}{q} \tanh(q d)$ we perform the integral numerically, and find that for $d = 3 \ell_B$, $J_c \sim 3.24$.
This is much larger than the region explored in our work ($J < 1$), indicating the attractive pairing is a collective effect.

\section{Phase diagram for other dopings}
\label{App:doping}
\begin{figure}
    \begin{subfigure}[b]{0.38\textwidth}
    \includegraphics[width=\textwidth]{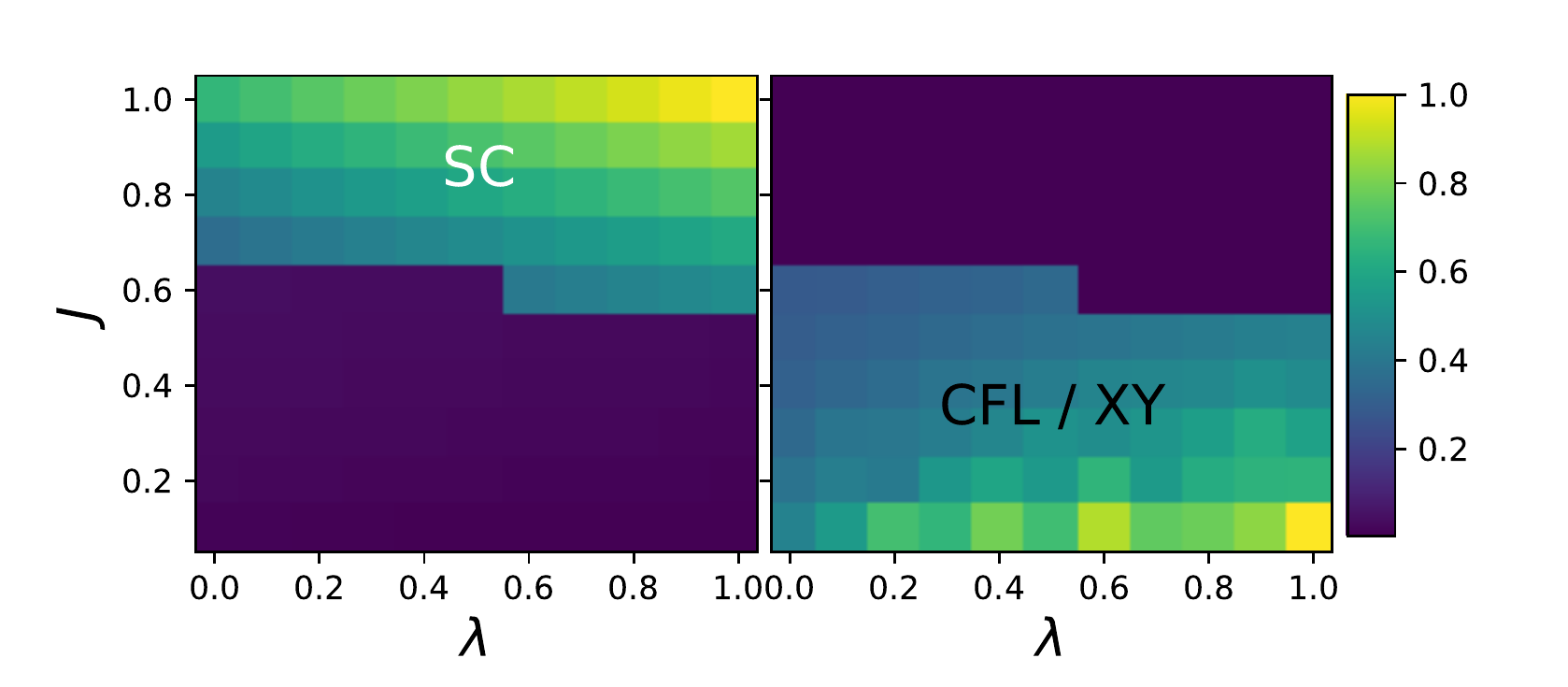}
    \caption{}
    \end{subfigure}
    \begin{subfigure}[b]{0.38\textwidth}
    \includegraphics[width=\textwidth]{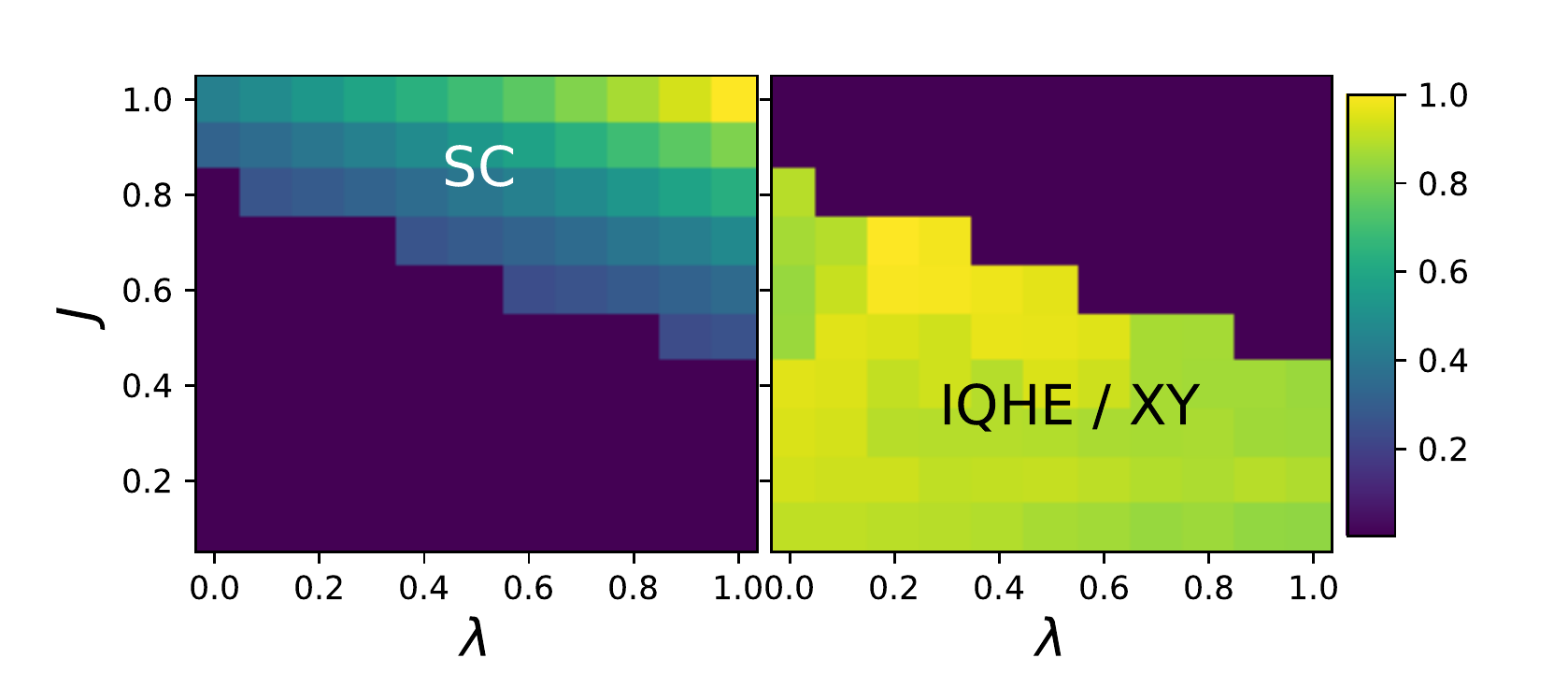}
    \caption{}
    \end{subfigure}
    \begin{subfigure}[b]{0.22\textwidth}
    \includegraphics[width=\textwidth]{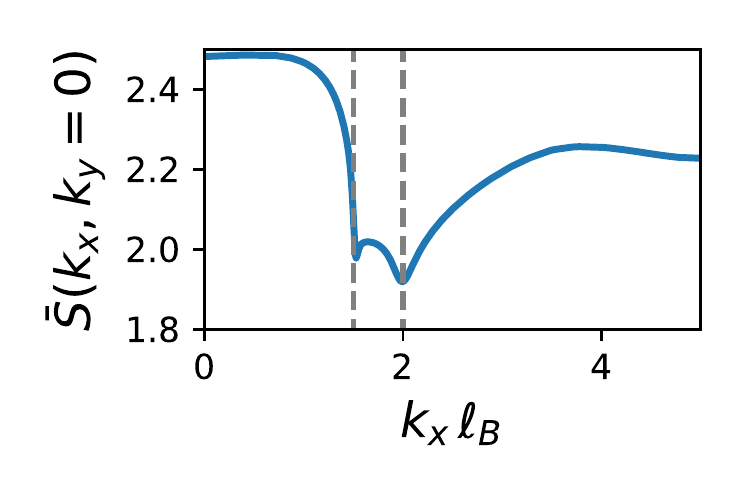}
    \caption{}
    \end{subfigure}
    \caption{(a) Phase diagram for doping $\delta = \frac{1}{2}$ calculated at $L_y = 10 \ell_B$. (b) Phase diagram for doping $\delta = 1$ calculated at $L_y = 10 \ell_B$. (c) Guiding center density-density structure factor in the layer-polarized phase ($J = 0.5, \lambda = 0.1$) at doping $\delta = \frac{1}{2}, L_y = 10 \ell_B$.
    The structure factor shows two singularities at wave-vectors $\ell_B k_x = 1.5, 2$. These values are exactly  consistent with the $2 k_F$ back scattering processes of the $\nu = \frac{1}{2}$ composite Fermi liquid, as discussed in Ref.~\onlinecite{geraedts2016half}.}
    \label{fig:PD_other_doping}
\end{figure}

Here we discuss the phase diagram for two other representative dopings: $\nu = 2 + \frac{1}{2}$, and $\nu = 2 + 1$, see Fig.~\ref{fig:PD_other_doping}.
As before, there are two phases: for large-$J$, the state is a layer-unpolarized SC, and for small-$J$, the doping $\delta$ layer polarizes on top of an XY-order. 
The large-$J$ SC region has the same properties as demonstrated for $\delta = \frac{1}{4}$,  e.g. the same pairing symmetry and an exponent $\eta_{\textrm{SC}} \sim L_y^{-1}$, so we will not discuss it further. We note that for the same $J, \lambda, L_y$, we find that  $\eta_{\textrm{SC}}$ decreases with the doping $\delta$. Presumably this is because the superfluid density $\rho_{\textrm{SC}}$, and the hence phase stiffness, increases with $\delta$.

The small-$J$ layer polarized phases ($\nu_+ = 1 + \delta, \nu_- = 1$) are more complex, but as we'll see they map onto a very familiar scenario:
fractional filling of a Landau level. 
In these phases, the  $\nu_- = 1$ component is essentially inert, and (in 2D)   polarizes along an isospin axis in the XY plane; for concreteness, let's say $\ket{-, \rightarrow}$. 
Due to the anti-ferromagnetic interlayer interaction, the electrons in layer $+$ then  effectively see an isospin Zeeman field of the form  $(J + \lambda) \eta^x$.
So, from the point of view of layer $+$, the problem is  qualitatively identical to a spinful Landau level at density $\nu_+ = 1 + \delta$ in the presence of a comparatively large Zeeman field $E_Z \sim J + \lambda$. 
The resulting phase diagram is well known. \cite{girvinLesHouches}
At $\nu_+ = 1$, the electrons spin-polarize into $\ket{+, \leftarrow}$: this is just the $\nu = 2$ XY-order. For small dopings $\nu_+ = 1 + \delta$, charge enters as either electrons or small charge-$1e$ skyrmions, forming CDWs such as Wigner crystals and or various bubble phases. This is the behavior found at $\delta = \frac{1}{4}$. As $\delta$ increases, it becomes favorable for the CDW to melt and give rise to various fractional quantum Hall states.
At $\delta = \frac{1}{2}$, for example, we find that and the $\nu_+ = 1 + \frac{1}{2}$ electrons form a composite Fermi liquid state! (Fig.~\ref{fig:PD_other_doping}) 
Finally, at $\delta = 1$, the $+$ layer is filled and becomes inert.

As can be seen, the phase boundary $J_c(\lambda, \delta)$ is doping dependent. This is expected. For small $\delta$, the energetics are dominated by $\Delta_{\textrm{pair}}$, which determines whether charge enters as electrons or layer-unpolarized charge-2e skyrmions. 
In this limit, $J_c \sim J_\ast$, as we found at $\delta = \frac{1}{4}$. 
As the doping $\delta$ increases, however, the energy of the SC and layer-polarized phases become sensitive to the interactions between the doped charges.
We see that for \textit{small} $\lambda$ this causes $J_c$ to increase with $\delta$, disfavoring the SC. 
This is presumably because for small $\lambda$ the charge-2e skyrmions are very large, while the charge-1e electrons are small, so the interaction energy increases with doping more rapidly in the SC phase.
In contrast, for large $\lambda$ we see that $J_c$ actually decreases with $\delta$ (albeit modestly), favoring the SC!

For small $\lambda$, our finding that $\frac{\partial J_c}{\partial \delta} < 0$  has an appealing consequence: it naturally leads to a superconducting ``dome'' as a function of the doping $\delta$.
For small $\delta$, the SC has a low $T_c$ because of the low superfluid weight, so as $\delta$ increases we expect that $T_c$ will at first increase (this is the usual density dependence of the BEC transition, though here the transition is BKT).
For large $\delta$, however, the system will eventually cross the $J_c$ boundary and the SC will be destroyed in favor of the symmetry-breaking layer polarized state. 
Depending on the system's precise location in the $(J, \lambda)$ plane, this leads to a situation where a SC dome emerges from $\nu = 2$, but then at some critical $\delta_\ast$ the SC is destroyed, evolving into the correlated insulator at e.g. $\nu = 3$. 
This scenario is reminiscent MATBG samples which show an insulator at neutrality \cite{efetov}.

\section{Bosonized description of the qXY/SC coexistence phase. \label{app:qXYbosonization}}

In the main text we claimed that the finite wave-vector $q_\ast$ of the qXY phase is consistent with the charges in the SC entering as meron-pairs, each of which binds a $\pi$-kink in the XY order parameter $\theta_{\textrm{XY}}$.
However, the reader may object that if the SC is a condensate of meron pairs, then this effect should actually destroy the XY order. 
However, because of the finite cylinder circumference, the system is an \textit{algebraic} SC and the variance in the number of pairs in a region grows only logarithmically, $\langle (\int^x_0\rho(x') dx')^2 \rangle_{\textrm{con}} \propto \log(x)$.
This can be used to infer the behavior of $\theta_{\textrm{XY}}$ using bosonization. So in this section we present a bosonized description of the qXY phase in which superconductivity coexists with finite-$q$ XY order and confirm the form of Eq.\eqref{eq:qXY}.

Let $\rho(x)$ denote the linear number density of meron pairs along the length of the cylinder at $x$, with 1D charge density $2 e \rho$.
We define slowly varying bosonic fields $\phi$ and $\tilde{\theta}$ which are related to the SC / XY order parameters via $\Delta(x) = \sqrt{\rho} e^{i \phi(x)}$ and $\theta_{\textrm{XY}}(x) = \tilde{\theta}(x) + \pi \int^x \rho(x') dx'$.
The latter expression realizes the constraint that the XY order parameter jumps by $\pi$ across each meron pair.
In the coexistence phase, both $\phi$ and $\tilde{\theta}$ are governed by  quadratic fluctuations which we assume (at long distances) decouple, with Luttinger parameters $K_{\textrm{SC}}, K_{\textrm{XY}}$.
We will show that  the resulting leading singularities in the SC and XY correlations take the form
\begin{align}
\langle \Delta^\dagger(x) \Delta(0) \rangle & \sim x^{- \frac{1}{2 K_{\textrm{SC}}}} + \cdots \\
\langle e^{i \theta_{\textrm{XY}}(x)} e^{-i \theta_{\textrm{XY}}(0)} \rangle &\sim x^{- \frac{1}{2 K_{\textrm{XY}}}} \langle e^{i \pi \int^x_0 \rho(x') dx'} \rangle \\
& \sim x^{- \frac{1}{2 K_{\textrm{XY}}} - \frac{\pi^2 K_{\textrm{SC}}}{2} }   \cos(\pi \rho_0 x) + \cdots
\end{align}
where $\rho_0$ is the average charge density and we have neglected subleading power laws.
As observed, the leading singularity shifts to finite $q_\ast = \pi \rho_0$.
As $J$ increases, the SC become stronger ($K_{\textrm{SC}}$ increases), and presumably the XY order is destroyed at a BKT transition.
This is consistent with the absence of XY order for large-$J$, where the central charge flows from $c=2 \to 1$.
In addition, because $K_{\textrm{SC}}$ increases linearly with the circumference $L_y$, this  causes the width of the qXY to shrink with circumference, so the qXY order does not survive in the 2D limit.

Because $\tilde{\theta}$ and $\phi$ are assumed to decouple in the IR, it  will be sufficient to compute $\langle e^{i \pi \int^x_0 \rho(x') dx'} \rangle$ in the SC phase. To do so, we follow the bosonization conventions of Ref.~\onlinecite{giamarchi2006strong} by introducing a phase field $\varphi_{l}$,
\begin{align}
\int^x_{-\infty} \rho(x') dx' &= \floor{\varphi_{l}(x) / 2 \pi}  \\  
e^{i \pi \int^x_{-\infty} \rho(x') dx' } &= e^{i \pi \floor{\varphi_{l}(x) / 2 \pi} } = \sum_{p \in \textrm{odd}} \frac{2}{i \pi p} e^{ i p \varphi_l(x) / 2}
\end{align}
where $\floor{x}$ is the floor function.
Expanding $\varphi_l(x) = 2 \pi \rho_0  x - 2 \varphi(x)$, we have
\begin{align}
\langle e^{i \pi \int^x_{-\infty} \rho(x') dx' } \rangle &=  \sum_{p, q \in \textrm{odds}} \frac{4}{\pi^2 p q}  e^{ i \pi p\rho_0  x}  \langle e^{i p  \varphi(x)}  e^{ - i q \varphi(0) }\rangle  \\
&=  \sum_{p \in \textrm{odds}} \frac{4}{\pi^2 p^2}  e^{ i \pi p \rho_0  x}  \langle e^{i p  \varphi(x)}  e^{ - i p \varphi(0) }\rangle \\
&=  \sum_{p \geq 0 \in \textrm{odds}} \frac{8 \cos( \pi p \rho_0  x)}{\pi^2 p^2} \frac{1}{x^{ K_{\textrm{SC}} p^2 \pi^2 /2 }} \approx \frac{8}{\pi^2} \frac{\cos( \pi \rho_0  x)}{x^{ K_{\textrm{SC}} \pi^2 /2 }} + \cdots
\end{align}
Note that in these conventions $\langle \Delta^\dagger(x) \Delta(0)\rangle \propto x^{-\frac{1}{2 K_{\textrm{SC}}}}$. The desired form then follows.

\section{Non-linear sigma model}
\label{App:NLSM}
In this section, we review and elaborate on aspects of the classical non-linear sigma model (NL$\sigma$M), including the critical $J_*(\lambda)$ required for pairing near half-filling, and its asymmetry between easy-plane ($\lambda > 0$) and easy-axis scenario $(\lambda < 0)$.

\begin{figure}
     \centering
     \begin{subfigure}[b]{0.48\textwidth}
         \centering
         \includegraphics[width=\textwidth]{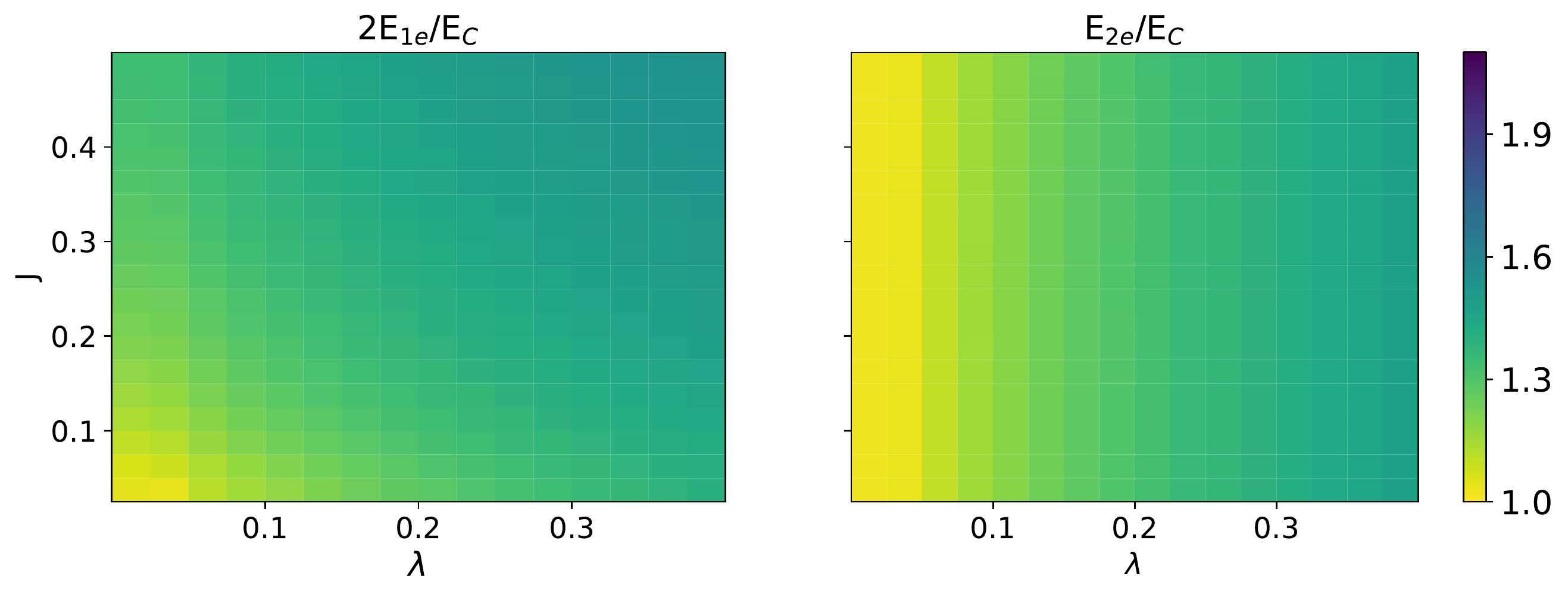}
         \caption{NL$\sigma$M}
         \label{fig:Ens_NLSM}
     \end{subfigure}  
     \hfill     
     \begin{subfigure}[b]{0.48\textwidth}
         \centering
         \includegraphics[width=\textwidth]{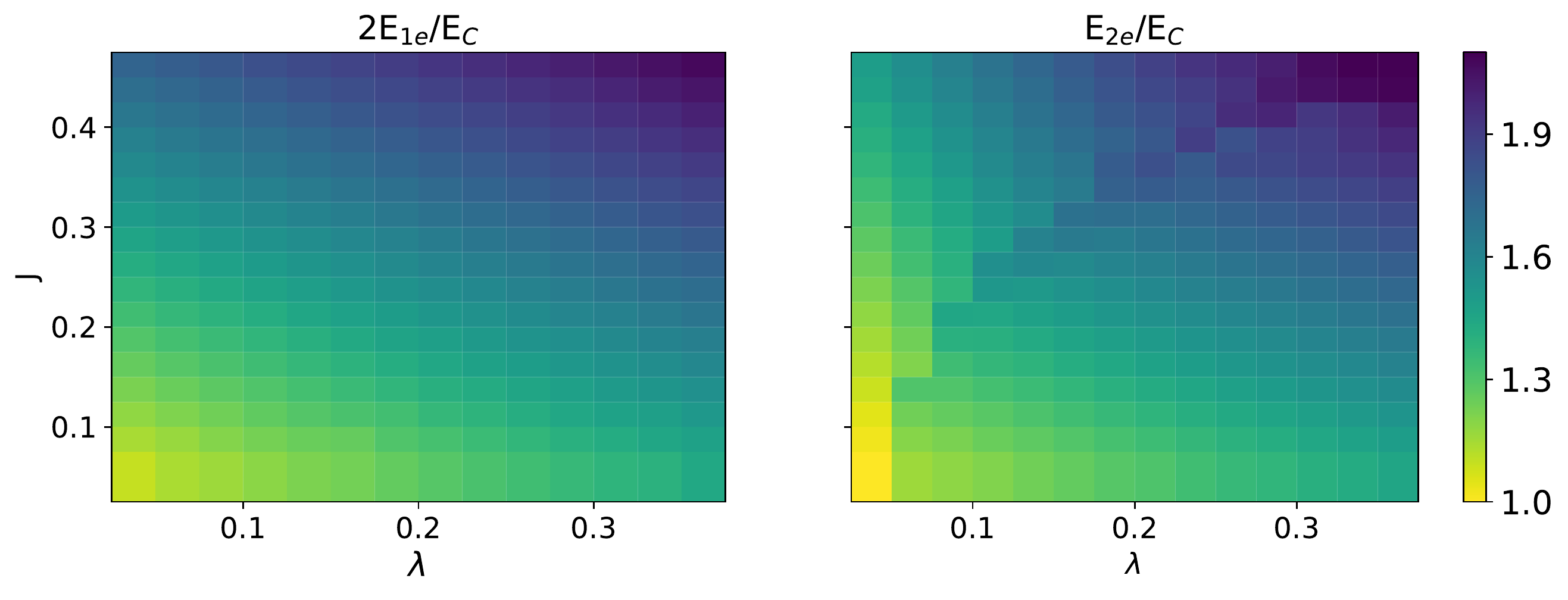}
         \caption{DMRG}
         \label{fig:Ens_DMRG}
     \end{subfigure}
        \caption{Energy of charge $e$ and charge $2e$ excitations, evaluated numerically. Note that $E_{2e}$ is consistently higher in DMRG than NL$\sigma$M; we attribute this to quantum fluctuations.}
        \label{fig:EnComp}
\end{figure}

We start by recalling the NL$\sigma$M partition function $Z = e^{-S}$ for coupled (iso)spin-ful lowest Landau levels in opposite magnetic fields, where the action given in imaginary time by $S = \int_0^\beta d\tau \int d^2\r \, \mathcal{L}[\n_+,\n_-]$ (Eq.~(\ref{eq:NLSM2}) in the main text with the identification $A_M = 2 \pi \ell_B^2$): 
\begin{align}
\mathcal{L}[\n_+,\n_-] = \sum_{\gamma = \pm} \left[ \frac{i}{2A_M} \int_0^1 du \, \n_\gamma \cdot (\partial_\tau \n_\gamma \times \partial_u \n_\gamma) + \frac{g}{2} (\nabla \n_\gamma)^2 \right] + \frac{ J_i  E_C}{2\pi A_M} (\n_+^i - \n_-^i)^2 + \frac{1}{2} \int d\r^\prime \rho(\r) V_c(\r - \r^\prime) \rho(\r^\prime), \nonumber \\
\text{ where }  \rho(\r) = \sum_{\gamma=\pm}\rho_\gamma(\r) = \sum_{\gamma = \pm} \frac{\gamma e}{4\pi} \n_\gamma \cdot (\partial_x \n_\gamma \times \partial_y \n_\gamma), ~ E_C = \frac{e^2}{4 \pi \epsilon \ell_B} \text{ and } A_M = 2 \pi \ell_B^2
\label{eq:NLSMapp}
\end{align}
The first term in Eq.~(\ref{eq:NLSMapp}) is the standard Berry's phase term for an isospin-half field\cite{girvinLesHouches,sachdev_2011}. The isospin-stiffness $g$ can be calculated in terms of the Coulomb energy scale $E_C = e^2/(4 \pi \epsilon \ell_B)$, for dual-gate-screened Coulomb potential of the form $V_C(\mathbf{q}) = V_C(q) = \frac{e^2}{2 \epsilon  q} \tanh(q d)$, as follows:
\begin{align}
g &= \frac{\ell_B^2}{32 \pi^2} \int_0^\infty dq\, q^3\, V_C(q)  e^{-(q \ell_B)^2/2} =  \frac{E_C}{16 \pi} \int_0^\infty dy \, \tanh\left(\frac{y d}{\ell_B}\right) y^2 e^{-y^2/2}     
\end{align}
For $d = 3\ell_B$, we find that $g \approx 0.99 g_0$, where $g_0 = E_C/(16\sqrt{2 \pi})$ is the value of isospin stiffness for unscreened Coulomb \cite{Sondhi}. For the numerics, we use this value of stiffness at different $J$ and $\lambda$ to extract the energy of charge $e$ and charge $2e$ excitations by minimizing the classical Hamiltonian on a $21 \ell_B \times 21 \ell_B$ square grid, with $\ell_B = 19$ units of grid spacing. The results for energetics are plotted in Fig.~\ref{fig:Ens_NLSM} and the relevant pairing energy $\Delta_{pair} = 2 E_{1e} - E_{2e}$ is shown in Fig.~\ref{fig:Excitations}a in the main text. We indeed find that pairing if favored at low anisotropy $\lambda$ and large $J$, which we can understand by using simple analytical calculations for the skyrmion energetics provided we neglect screening.   

\begin{figure}
     \centering
     \begin{subfigure}[b]{0.47\textwidth}
         \centering
     \includegraphics[width=\textwidth]{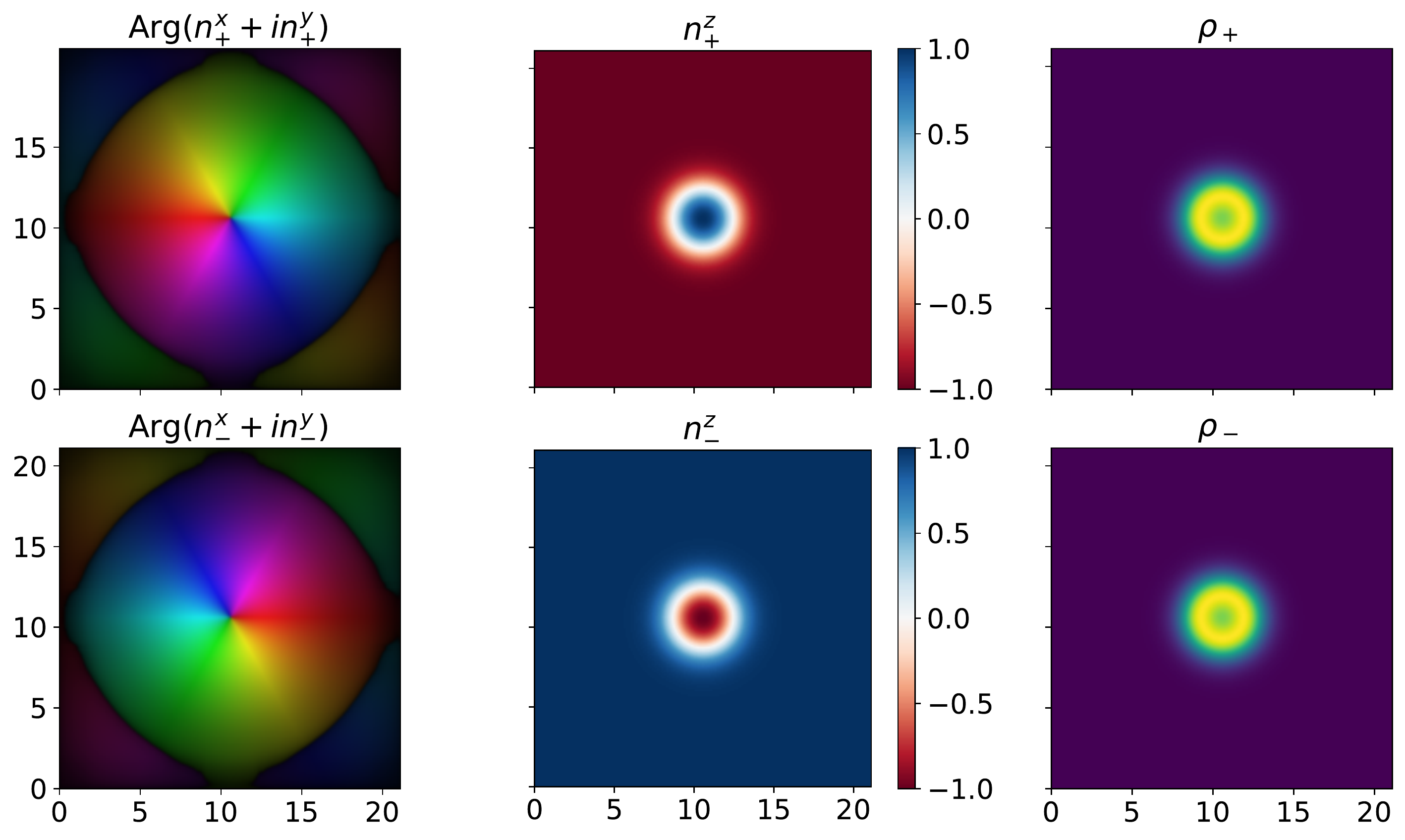}
         \caption{Easy-axis: $J = 0.5, \lambda = -0.3$}
         \label{fig:EAns}
     \end{subfigure}  
     \hfill     
     \begin{subfigure}[b]{0.47\textwidth}
         \centering
         \includegraphics[width=\textwidth]{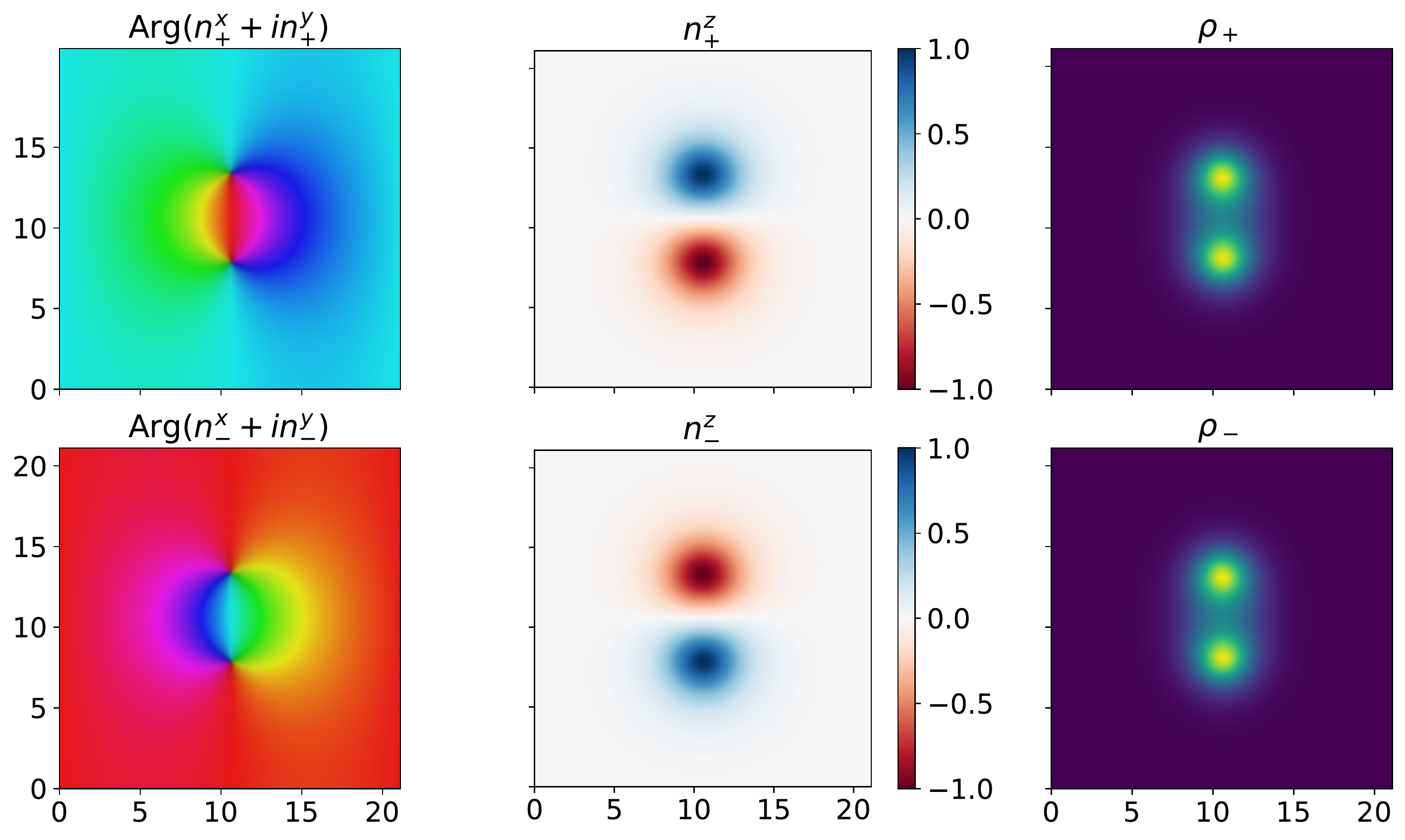}
         \caption{Easy-plane: $J = 0.5, \lambda = 0.3$}
         \label{fig:EPns}
     \end{subfigure}
        \caption{Distribution of spin and charge density in the two layers ($\gamma = \pm$) for $2e$ charged excitations obtained by numerics on the classical NL$\sigma$M. While the spin-density is always locally antiferromagnetic, the charge density is radially symmetric for easy-axis skyrmions, but splits into two merons for easy-plane skyrmions at large $\lambda/\rho_s$.}
        \label{fig:merons}
\end{figure}

Since the easy-plane case has been discussed in detail in Ref.~\onlinecite{Eslam}, here we focus on the easy-axis case, and show how the phase-boundary $J_{\ast}(\lambda)$ can be well-captured by a variational texture with a single tunable parameter, the radius $R$ of the skyrmion. We first consider the following ansatz for a charge $e$ skyrmion in a single layer (and neglect possible weak back-reaction from the opposite layer).
\begin{align}
\n_+(\r) =& (\sin \Theta \cos \Phi, \sin \Theta \sin \Phi, \cos \Theta), \text{ with } \Theta(\r) = \theta(r) = 2 \arcsin(e^{-r/2R}) \text{ and } \Phi(\r) = \phi,\nonumber \\
\n_-(\r) =& (0,0,-1)
\end{align}
 The total energy of such a texture (for unscreened Coulomb interaction) is given by the sum of its elastic, exchange (Zeeman) and Coulomb charging energy:
\begin{equation}
E_{1e}(R) = 4.4 \pi g + \frac{4 E_C (J + |\lambda|)R^2}{A_M} +  \frac{e^2}{16 \epsilon R} 
\label{eq:En1skR}
\end{equation}
The optimal size (and consequently energy) is controlled by the competition between the Coulomb charging energy and effective Zeeman energy penalty due to loss of antiferromagnetic exchange with the opposite layer.
\begin{equation}
R_{opt} = \left( \frac{\pi^2 }{16 (J + |\lambda|)} \right)^{1/3} \ell_B, \text{ and } E_{1e}(R_{opt}) = 4.4 \pi g + 3 \left( \frac{\pi (J + |\lambda|) }{4} \right)^{1/3}
\end{equation}. 

For the charge $2e$-skyrmion, we consider a locally antiferromagnetic ansatz of the form:
\begin{align}
\n_\pm(\r) =& (\sin \Theta_\pm \cos \Phi_\pm, \sin \Theta_\pm \sin \Phi_\pm, \cos \Theta_\pm), \text{ with } \Theta_+(\r) = \theta_+(r) = 2 \arcsin(e^{-r/2R}), \Phi_+(\r) = \phi \nonumber \\
\Theta_-(\r) =& \pi - \Theta_+(\r), \Phi_-(\r) = \phi + \pi, \iff \n_-(\r) = - \n_+(\r) 
\end{align}
The total excitation energy of this texture is independent of $J$ as local antiferromagnetism is perfectly respected, and is given by:
\begin{equation}
E_{2e}(R) = 8.8 \pi g + \frac{12 E_C |\lambda| R^2}{A_M} +  \frac{e^2}{4 \epsilon R}   
\end{equation}
The optimal size is therefore determined by the competition between Coulomb energy and anisotropy, leading to 
\begin{equation}
R_{opt} = \left( \frac{\pi^2 }{12 |\lambda|} \right)^{1/3} \ell_B, \text{  and } E_{2e}(R_{opt}) = 8.8 \pi g + 3 E_C(12 \pi |\lambda|)^{1/3}.    
\end{equation} 
From this, we determine the minimum exchange $J$ for a given anisotropy $\lambda$, beyond which charge $2e$ excitations become lower in energy: $
2 E_{1e} \geq E_{2e} \implies J \geq 5 |\lambda|$. Thus, we see that $J_*(\lambda) = 5|\lambda|$ for our ansatz. In particular, our calculation implies that $J_*(\lambda \to 0) \to 0$; in this limit $R_{opt}$ for the $2e$ skyrmion diverges and it completely evades any Coulomb energy cost. For screened Coulomb interaction, we expect the critical $J_*(\lambda)$ to be lower. This is confirmed by our numerics with screened Coulomb interaction having $d = 3 \ell_B$, where we find that the dotted purple line on the easy axis side of Fig.~\ref{fig:Excitations}a is approximately linear with $J_*(|\lambda|) \approx 3.5 |\lambda|$.

An analogous computation in the easy-axis case \cite{Eslam} leads to a smaller slope for critical $J_{\ast}(\lambda) = 2 \lambda$, indicating that pair-formation is favorable in the easy-axis case. Roughly speaking, within this variational ansatz this is because canting of isospin in the direction normal to the ordering vector in the easy-plane scenario does not cost additional anisotropy energy, in contrast to the easy-axis case where any canting away from the easy-axis incurs an additional anistropy energy cost. However, there is a more significant reason which is not captured by such an ansatz; for small stiffness $g/E_C$, it is more favorable for the charge $2e$ object to deform into a topologically equivalent texture consisting of two charge $e$ merons confined by an elastic binding force, while still maintaining perfect local antiferromagnetism. This is evidenced by the plot of charge density in Fig.~\ref{fig:merons}, clearly showing the separation of the charge density into merons for the easy-plane case and a radially symmetric distribution for charge density in the easy axis case, for the same value of $(J,|\lambda|)$. Indeed, an analytical calculation \cite{Eslam} shows that $J_\ast(\lambda) \to 0$ in the limit of small isospin stiffness relative to the anisotropy ($g/\lambda \to 0$).

\section{Details of segment DMRG and comparison with NL$\sigma$M}
\label{App:SegDMRG}
In this section, we first elaborate on the details of extracting the energy of charge-$2e$ and charge-$e$ excitations above the antiferromagnetic insulating ground state at $\nu = 2$. Later, we discuss quantitative differences between the quantum and classical energetics and discuss quantum fluctuations as a possible origin.

As discussed in the main text, we consider two classes of excitations: either a single electron in one layer (1e), or two electrons with one in each layer (2e). The minimal excitation energy of each such excitation relative to the ground state at $\nu = 2$ is extracted as follows. For matrix product state (MPS) fixed bond dimension $\chi$ and cylinder circumference $L_y$, we allow the MPS representation of the quantum wave-function to differ from the ground state on an axial segment that spans $N_{span}$ Landau level orbitals per spin per layer (in the Landau gauge).
Within this variational space, DMRG is used to find the minimum energy excitation with fixed quantum numbers for charge (e or 2e), spin and layer polarization of the excited state wave-function relative to the ground state. For a given charge and layer polarization, the spin quantum number corresponding to minimum excitation energy is chosen. Finally, appropriate extrapolations as functions of $\chi$, $N_{span}$ and $L_y$ are performed to obtain $E_{e}$ or $E_{2e}$ in the thermodynamic limit.

\begin{figure}
     \centering
     \begin{subfigure}[b]{0.9\textwidth}
         \centering
         \includegraphics[width=\textwidth]{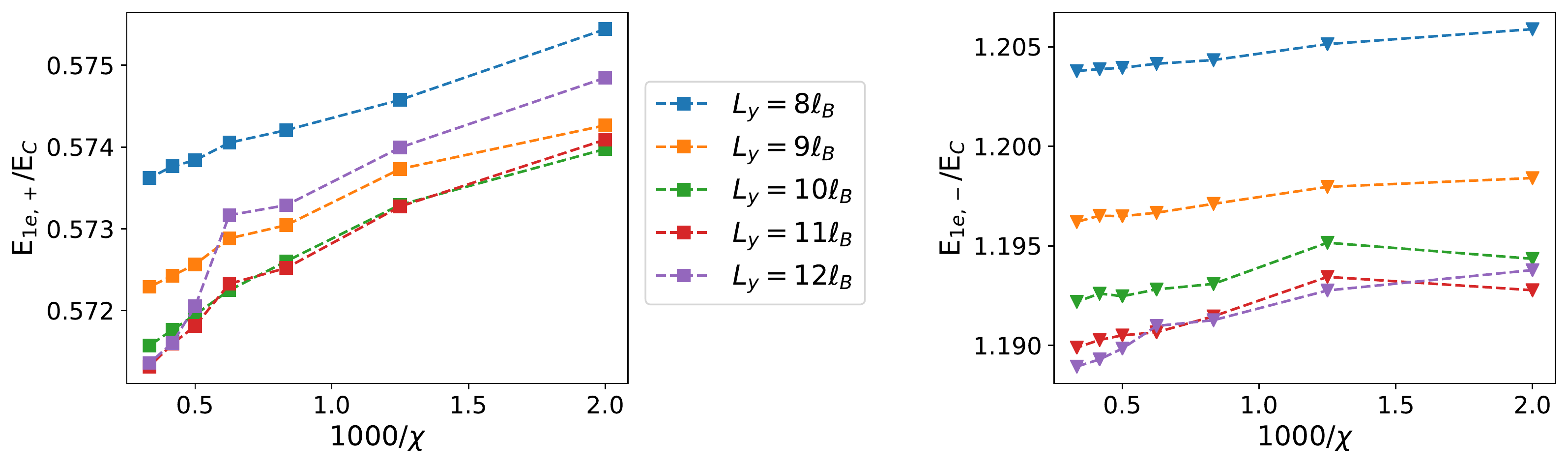}
         \caption{DMRG energy $E_{1e,\gamma}$ of charge e excitations in each layer ($\gamma = \pm 1$) as a function of bond dimension $\chi$ for different cylinder circumference $L_y$, showing that it is nearly independent of bond dimension for $\chi \gtrsim 2000$, but depends weakly on $L_y$ which necessitates extrapolation.}
         \label{fig:EneVsChi}
     \end{subfigure}  
     \hfill
     \begin{subfigure}[b]{0.9\textwidth}
         \centering
         \includegraphics[width=\textwidth]{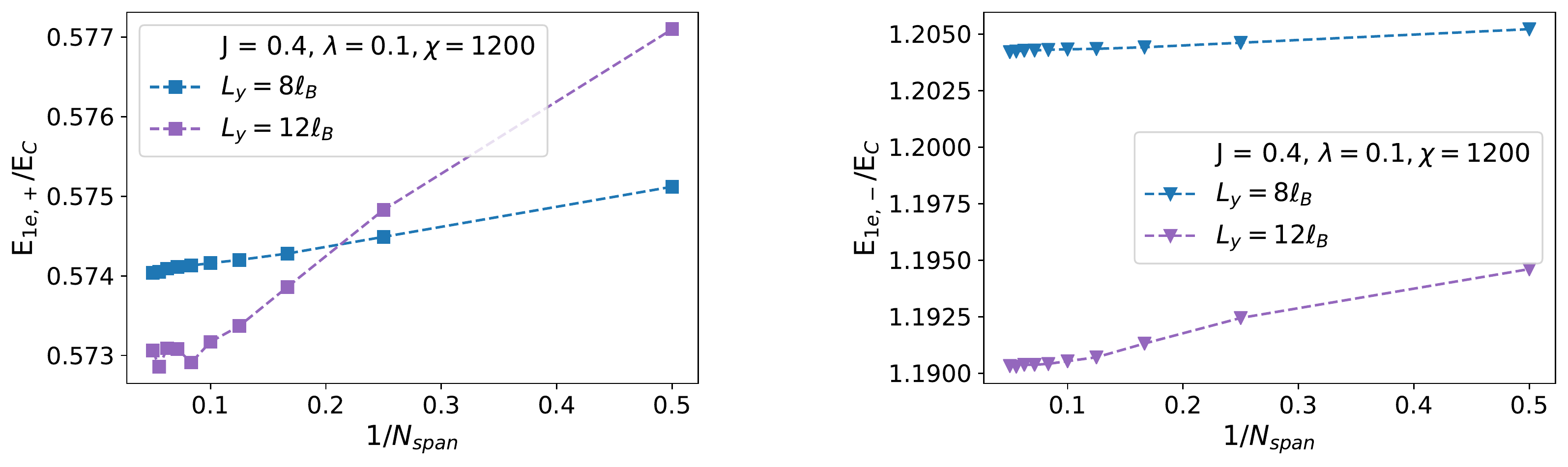}
         \caption{DMRG energy $E_{1e,\gamma}$ of charge e excitations in each layer ($\gamma = \pm 1$) as a function of span of axial segment, showing convergence is reached for $N_{span} \gtrsim 8$.}
         \label{fig:EneVsSpan}
     \end{subfigure}
     \hfill
     \begin{subfigure}[b]{0.9\textwidth}
         \centering
         \includegraphics[width=\textwidth]{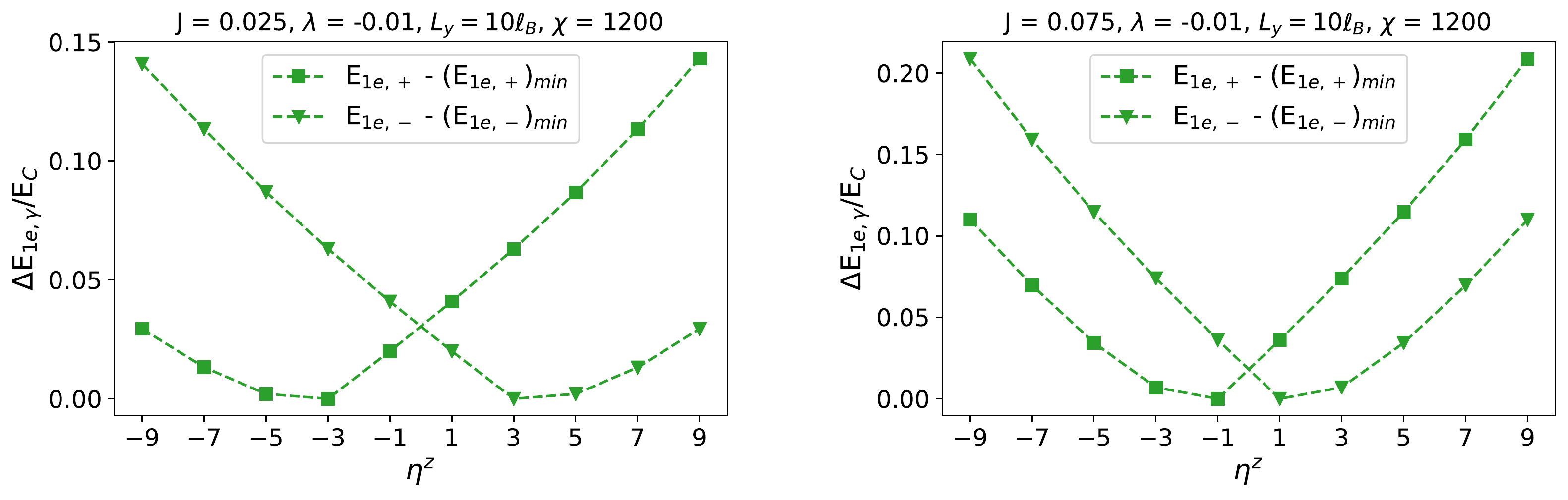}
         \caption{Energy of charge $e$ excitations as a function of spin quantum number $\eta^z$ in the easy axis scenario ($\lambda < 0$), with the minimum subtracted off to show the symmetric structure about $\eta^z = 0$. For larger effective Zeeman coupling to the opposite layer (proportional to $J + |\lambda|$) the energy minima occur at $\eta^z \pm 1$ for charges in opposite layers. For smaller effective Zeeman field, the minima shift to $\eta^a = \pm 3$, strongly indicating that charge $e$ excitations are actually topological skyrmions.}
         \label{fig:EneVsSpin}
    \end{subfigure}
        \caption{DMRG energy of charge $e$ excitations above the insulator at charge neutrality as functions of $\chi$, $L_y$, $N_{span}$ and $\eta^z$}
        \label{fig:EneData}
\end{figure}

\subsection{Charge-$e$ excitations} 
\label{appss:e}
\begin{figure}
     \centering
     \begin{subfigure}[b]{0.55\textwidth}
         \centering
         \includegraphics[width=\textwidth]{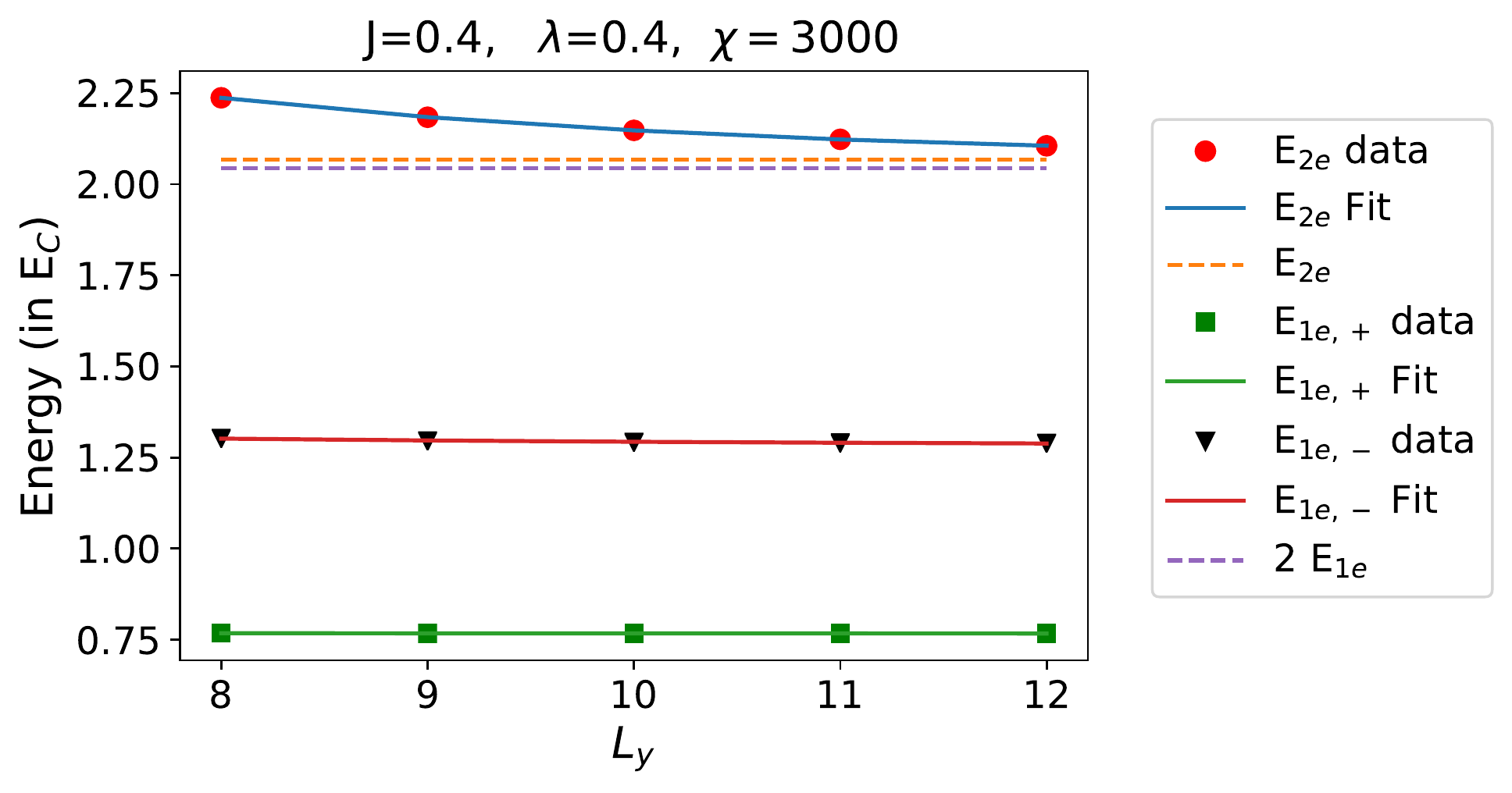}
         \caption{Pair formation disfavored at large anisotropy}
         \label{fig:Enfit1}
     \end{subfigure}  
     \hfill     
     \begin{subfigure}[b]{0.40\textwidth}
         \centering
         \includegraphics[width=\textwidth]{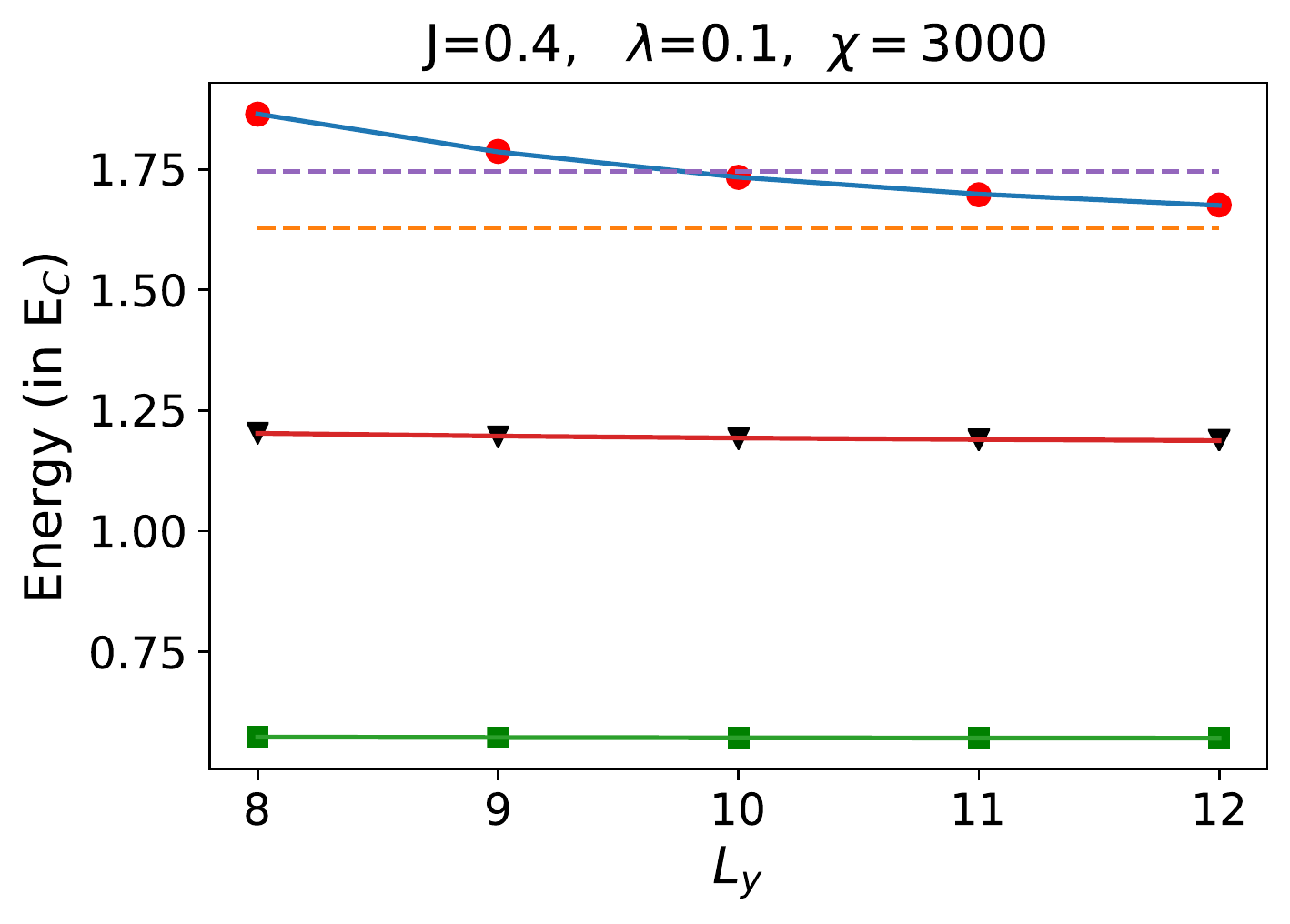}
         \caption{Pair formation favored at low anisotropy}
         \label{fig:Enfit2}
     \end{subfigure}
        \caption{Energy of charged excitations as a function of $L_y$, with best fits and extrapolated values in the thermodynamic limit. When the dotted yellow line ($E_{2e}(L_y \to \infty)$) lies below the dotted purple line ($2 E_{1e}(L_y \to \infty)$), pair formation is favored.}
        \label{fig:Enfit}
\end{figure}

We find that the charge $e$ excitation energy $E_{1e}$ (for either layer) does not depend much on $N_{span}$ or bond dimension $\chi$, and depends very weakly on the cylinder circumference $L_y$. Further, in the easy-plane antiferromagnet ($\lambda > 0$) where the ground state spontaneously breaks $\eta^z$, there is negligible dependence on the spin quantum number $\eta^z$ for $E_{1e}$ in either layer. Therefore, we work with fixed $\eta^z = 1$ in a regime of $N_{span}$ and $\chi$ where $E_{1e}$ has already converged as a function of segment length and bond dimension, and extract $E_{1e}$ in the thermodynamic limit $L_y \to \infty$ by extrapolation. Excellent fits are obtained for $E_{1e}(L_y) = E_{1e} + a_{1}/L_y + b_{1}/L_y^2$ (representative fits are shown in Fig.~\ref{fig:Enfit}, indicating that the spin-half charge $e$ excitations are well-localized within the screening length $d = 3 \ell_B$; consistent with our picture that the excitations are simply localized electrons. The extrapolated $2E_{1e} = \sum_{\gamma = \pm} E_{1e,\gamma}$ in $(J,\lambda)$ plane are shown in Fig.~\ref{fig:Ens_DMRG}.

Qualitatively similar behavior is observed for $E_{1e}$ in the easy-axis antiferromagnet ($\lambda < 0$). However, in this case the ground state conserves total $\eta^z$, and consequently $E_{1e}$ for each layer has a marked dependence on the spin $\eta^z$, which needs to be aligned anti-parallel to the spin of the opposite layer to gain energy. Further, we find that for small values of the effective Zeeman field from the opposite layer, i.e, $J + |\lambda| \lesssim 0.06$, the minimum energy charge $e$ excitations have $|\eta^z| = 3$, providing strong evidence that these excitations are topological skyrmions (see Fig.~\ref{fig:EneVsSpin}). This is also consistent with the predictions of the classical NL$\sigma$M, where the size of the skyrmion (and therefore its spin) is determined by the competition between Zeeman and Coulomb energy, and therefore decreases with increasing effective Zeeman field. 

\subsection{Charge-$2e$ excitations} 
\label{appss:2e}
\begin{figure}
     \centering
     \begin{subfigure}[b]{0.55\textwidth}
         \centering
         \includegraphics[width=\textwidth]{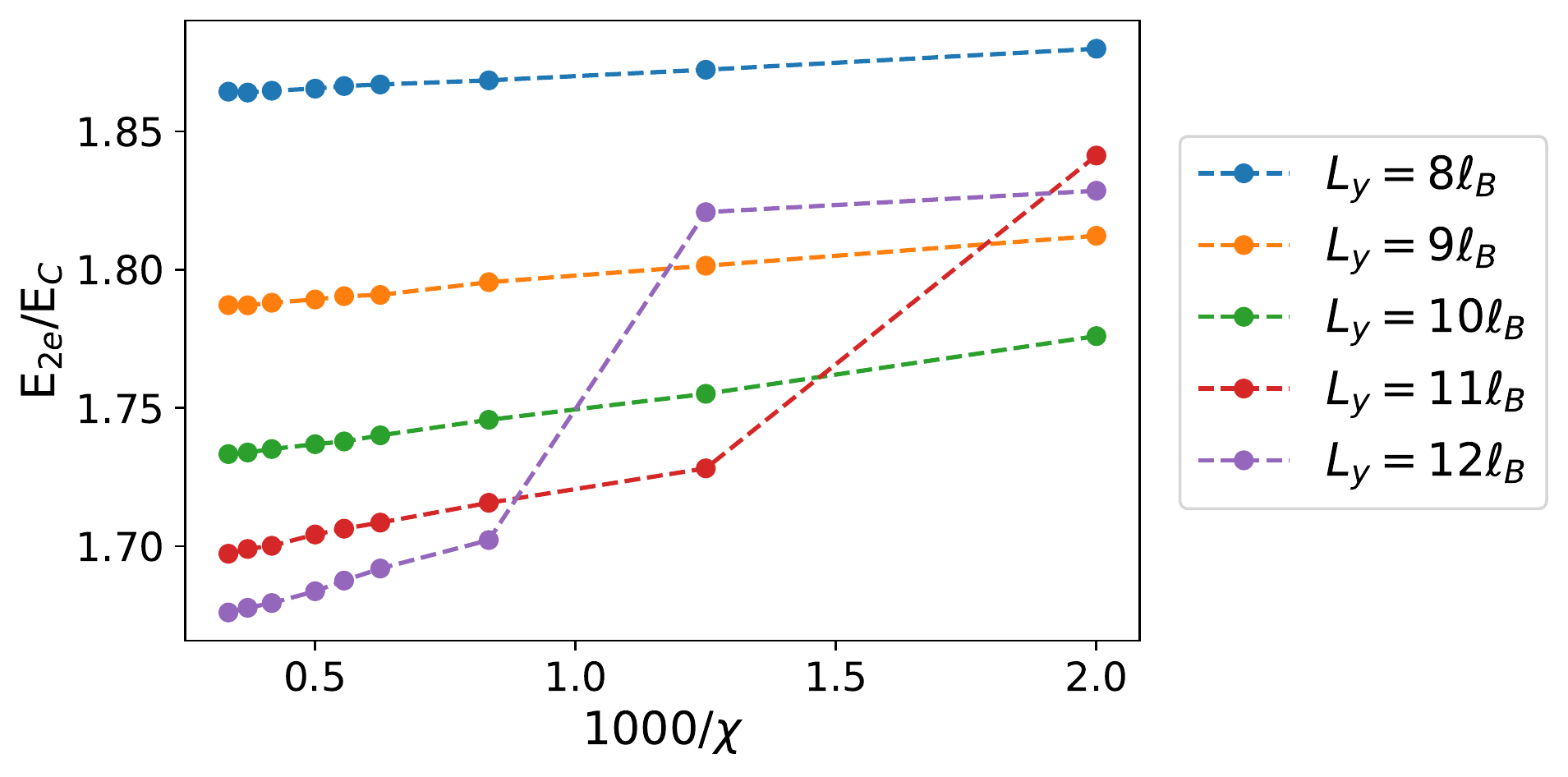}
         \caption{DMRG pair energy $E_{2e}$ as a function of bond dimension $\chi$ for different cylinder circumference $L_y$ at $N_{span} = 20$, showing that convergence is reached for $\chi \gtrsim 2000$, but extrapolation is required as a function of $L_y$.}
         \label{fig:En2eVsChi}
     \end{subfigure}  
     \hfill     
     \begin{subfigure}[b]{0.40\textwidth}
         \centering
         \includegraphics[width=\textwidth]{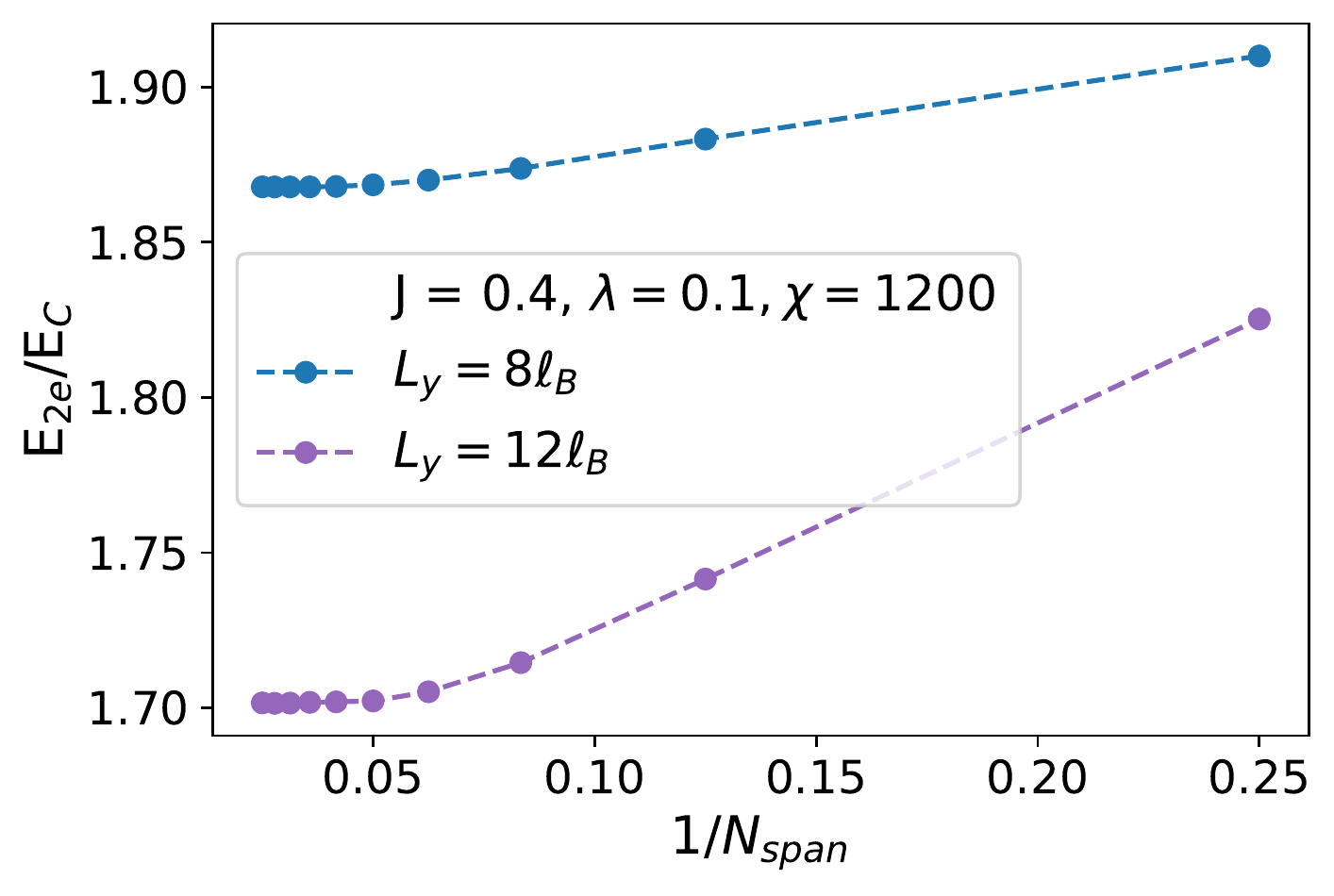}
         \caption{DMRG pair energy $E_{2e}$ as a function of span of axial segment at two $L_y$, showing convergence is reached for $N_{span} \gtrsim 20$.}
         \label{fig:En2eVsSpan}
     \end{subfigure}
        \caption{DMRG energy of charge $2e$ excitations above the insulator at charge neutrality as functions of $\chi$, $L_y$ and $N_{span}$}
        \label{fig:En2eData}
\end{figure}

Next, we turn to the energetics of charge $2e$ excitations above the ground state. Typically, the dependence on bond-dimension is negligible beyond a certain minimum $\chi$ that depends on J and $\lambda$, as shown in Fig.~\ref{fig:En2eVsChi}. However, the charge now prefers spread out over much larger length-scales, indicating the need for larger $N_{span}$ and $L_y$ to accurately extract $E_{2e}$ in the thermodynamic limit. Once again, we find that $E_{2e}$ converges rapidly beyond $N_{span} = 20$ (see Fig.~\ref{fig:En2eVsSpan}); therefore we fix $N_{span} = 20$ and extrapolate as a function of $L_y$. Excellent fits are obtained for $E_{2e}(L_y) = E_{2e} + a_{\ell} e^{- L_y/\ell_s}$ (for representative fits see Fig.~\ref{fig:Enfit}), indicating that the spin-zero charge $2e$ excitations are extended well-beyond within the screening length $d = 3 \ell_B$, and therefore the Coulomb energy goes down exponentially for $L_y \gg d$; this is further evidenced by noting that $\ell_s \approx 3\ell_B = d$ in our fits. This is consistent with our classical picture that the charge $2e$ excitations above the ground state are non-trivial topological textures which can completely avoid Coulomb repulsion by spreading out to a large size for small anisotropy. The extrapolated $E_{2e}$ in the $(J,\lambda)$ plane are shown in Fig.~\ref{fig:Ens_DMRG}.

\subsection{Effective mass} 
\label{appss:Mpair}
Having discussed the details of energetics at $k_y = 0$, we now elaborate on the extraction of dispersion relations at non-zero $k_y$. This is achieved by varying the cylinder circumference $L_y$ for the segment DMRG, allowing us to access momenta $k_{y,n} = 2 n \pi/L_y$ with $n \in \mathbb{Z}$ \cite{ZaletelEFS}. As discussed in the main text and shown in Fig.~(\ref{fig:dipole}), finite $k_y$ results in the $2e$ skyrmion splitting into two charge $e$ excitations in opposite layers, which move towards opposite edges of the cylinder. This can be seen by noting that the lowest Landau level wavefunction at momentum $k_y$ in the Landau gauge is peaked at $\langle x \rangle = \pm k_y \ell_B^2$, corresponding to Chern number $\pm 1$ respectively; a net momenta $k_y$ therefore results in a separation $\Delta x = k_y \ell_B^2$. This makes the $2e$ pair lose local antiferromagnetic exchange energy, which serves as binding glue, at large $k_y$, and the energetics is now dominated by Coulomb repulsion between the two charge $e$ excitations. Consequently, the dispersion becomes non-monotonic. In practice, we find that this physics takes over for $k_y \ell_B \gtrsim 1.5$. This, behavior, along with significant finite size effects, make it difficult to extract an effective mass.

\begin{figure}
    \centering
    \includegraphics[width =0.99\textwidth]{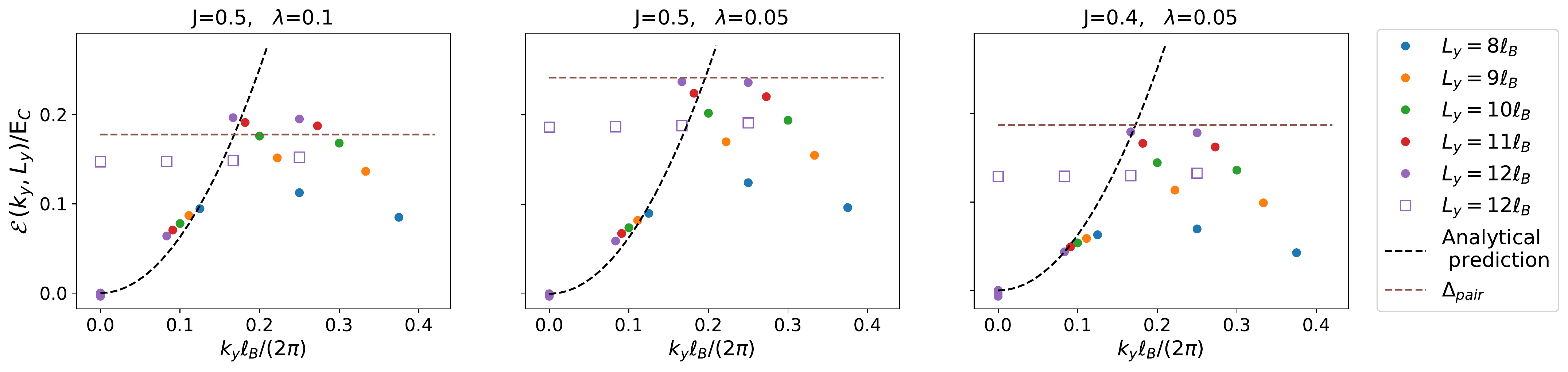}
    \caption{Dispersing $2e$ charged excitations (filled circles), and non-dispersing $1e$ charged excitations (empty squares) for different values of $(J,\lambda)$. The predicted classical isotropic dispersion compares reasonably well with the numerically computed dispersion at small $k_y \ell_B$, indicating that the effective mass is approximately independent of $\lambda$ (compare first two panels) and scales inversely with $J$ (compare last two panels) at small anisotropy.}
    \label{fig:MpairApp}
\end{figure}

Nevertheless, we can still try to compare the energy at small $k_y$ and small anisotropy $\lambda$, to the semiclassical dispersion expected from analytic calculations corresponding to $M_{pair} = \pi \hbar^2/(J_{p} A_M^2)$ \cite{Eslam} in the isotropic limit ($J_p$ being the antiferromagnetic coupling between opposite Chern sectors, the equivalent of \emph{layers} in Ref.~\onlinecite{Eslam}). In our convention, $A_M = 2 \pi \ell_B^2$ and $J_p A_M = E_C J/\pi$, therefore after appropriate conversion $M_{pair} = \pi \hbar^2/(2 E_C \ell_B^2 J) = \pi/(2J)$ in the units used in Figs.~\ref{fig:Excitations} and \ref{fig:MpairApp}. To eliminate finite size effects, we plot $\mathcal{E}_{2e}(k_y, L_y) \equiv E_{2e}(k_y, L_y) - E_{2e}(k_y = 0, L_y)$, and see that at small $k_y \ell_B$ the expected isotropic dispersion given by $k_y^2/(2 M_{pair})$ matches quite well. We further note from Fig.~\ref{fig:MpairApp} that while the effective mass roughly scales as $J^{-1}$ as predicted by the semiclassical calculations, the dependence on anisotropy $\lambda$ is quite weak. Finally, we comment that at large $k_y \ell_B \gtrsim 1.5$ and large $L_y$, the energy of the $2e$ excitation $E_{2e}(k_y) \to 2 E_{1e}$, as evidenced by $\mathcal{E}_{2e}(k_y) \to \Delta_{pair}$ in Fig.~\ref{fig:MpairApp}.

\subsection{DMRG vs NL$\sigma$M} 
\label{appss:compare}
\begin{figure}
    \centering
    \includegraphics[width=0.5\textwidth]{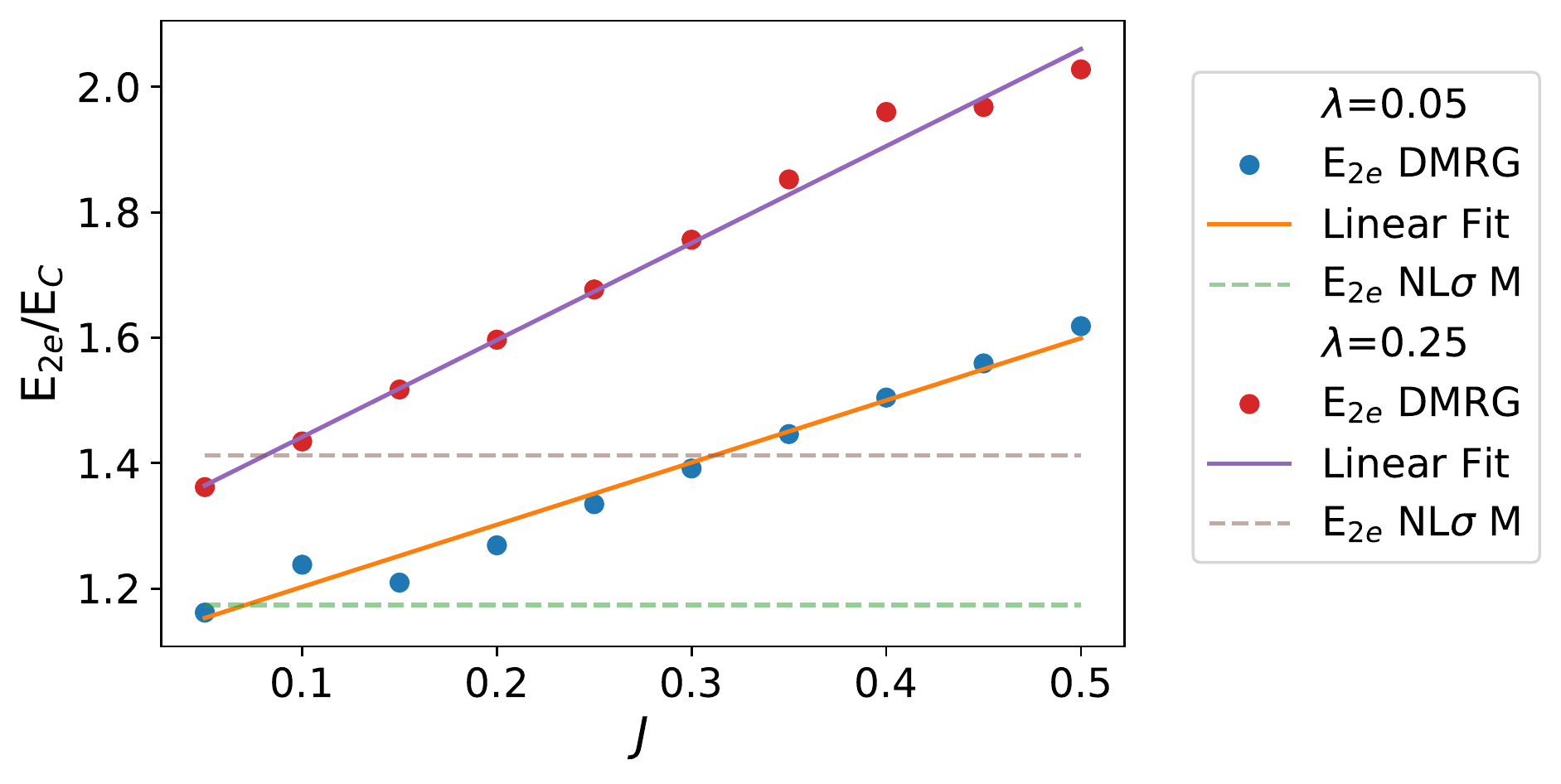}
    \caption{$E_{2e}$ increases linearly in $J$ for a given $\lambda$ in DMRG, while it remains constant in NL$\sigma$M due to perfect local antiferromagnetism. For small $J$, both approaches yield very similar $E_{2e}$, indicating that quadratic quantum fluctuations are almost entirely responsible for correction. Larger slope of the $E_{2e}(J)$ line for increasing anisotropy can be interpreted as increased quantum fluctuations due to decrease is size of $2e$ skyrmions with increasing anisotropy.}
    \label{fig:En2eVsJ}
\end{figure}

Although there is good semi-quantitative agreement between the classical NL$\sigma$M and the quantum energetics found via segment DMRG, there are some minor discrepancies. In particular, we find that the DMRG energy of the charge $2e$ excitation increases with $J$ at a fixed $\lambda$, although there is no dependence of $E_{2e}$ on $J$ in the classical picture due to perfect local antiferromagnetism between the layers. To resolve this, we first note that $E_{2e}$ is always found to be minimum at $\eta^z = 0$, corresponding to perfect antiferromagnetic alignment between the opposite layers (see Fig.~\ref{fig:EnComp}). Therefore, the additional contribution must come from quantum fluctuations, which we aim to quantify as a function of $J$. To this end, we proceed by integrating out the ferromagnetic modes in the quantum NL$\sigma$M action. The procedure closely resembles integrating out ferromagnetic modes for a two-dimensional collinear Heisenberg antiferromagnet \cite{Haldane83,sachdev_2011}; we decompose $\n_{\gamma}$ as follows:
\begin{equation}
\n_\gamma(\r) = \gamma \n(\r) \sqrt{1 - \m^2(\r)} + \m(\r), ~~~ |\m(\r)| \ll 1, ~ \n \cdot \n = 1 \text{ and } \n \cdot \m = 0
\end{equation}
Plugging this into the action in Eq.~(\ref{eq:NLSMapp}) and assuming slow variation in space so that we can neglect terms with two or more derivatives \text{and} two or more powers of $\m$ (i.e, O($k^2 m^2$) terms with $k$ being momenta), we arrive at the following coupled action for $\n$ and $\m$:
\begin{align}
\mathcal{L}[\n,\m] =& \frac{i}{A_M} \m \cdot (\n \times \partial_\tau \n) + g (\nabla \n)^2 + \frac{2E_C}{\pi A_M} \left[ \lambda (n^z)^2 + (J + \lambda \n_{xy}^2) \m_{xy}^2 + [J + \lambda(1 + \n_{xy}^2)] m_z^2 \right]  - \mu \rho(\r) \nonumber \\ & + \frac{1}{2} \int d\r^\prime \rho(\r) V_C(\r - \r^\prime) \rho(\r^\prime), \text{ where } \rho(\r) = \frac{2e}{4\pi} \n \cdot (\partial_x \n \times \partial_y \n)
\end{align}
At this point, the action is quadratic in $\m$, and we can integrate out $\m$ to find an effective action within the antiferromagnetic manifold $\n$.
\begin{align}
\mathcal{L}_{eff}[\n] =  \frac{1}{2 A_M^2} (\n \times \partial_\tau \n)_i [\mathcal{A}^{-1}]_{ij} (\n \times \partial_\tau \n)_j + g(\nabla \n)^2 + \frac{2 \lambda E_C}{\pi A_M} (n^z)^2 - \mu \rho(\r) + \frac{1}{2} \int d\r^\prime \rho(\r) V_C(\r - \r^\prime) \rho(\r^\prime), \nonumber \\
\text{ where } \mathcal{A} = \frac{4 E_C}{\pi A_M} \begin{pmatrix} (J + \lambda \n_{xy}^2) & 0 & 0 \\ 0 & (J + \lambda \n_{xy}^2) & 0 \\ 0 & 0 &  J + \lambda(1 + \n_{xy}^2) \end{pmatrix}
\end{align}
While the Hamiltonian corresponding to this effective action can be found via analytic continuation to real time followed by a Legendre transform, it is non-illuminating and cumbersome to write down. Since pairing is seen only for small anisotropy, it is instructive to consider the isotropic limit, in which case the effective Lagrangian reduces to (after analytic continuation to real time):
\begin{equation}
\mathcal{L}_{eff}[\n] = \frac{\pi}{8 J A_M E_C} (\partial_t \n)^2 - g(\nabla \n)^2 - \frac{2 \lambda E_C}{\pi A_M} (n^z)^2 + \mu \rho(\r) - \frac{1}{2} \int d\r^\prime \rho(\r) V_C(\r - \r^\prime) \rho(\r^\prime),
\end{equation}
In this limit, the conjugate momenta $\mathbf{L} = \left( \frac{\pi}{4 J A_M E_C}\right) \partial_t \n$, and the effective quantum Hamiltonian density is given by:
\begin{equation}
H_{eff}[\mathbf{L},\n] = \left( \frac{2 J A_M E_C}{\pi} \right) \mathbf{L}^2 +  g(\nabla \n)^2 + 2 \lambda (n^z)^2 - \mu \rho(\r) + \frac{1}{2} \int d\r^\prime \rho(\r) V_C(\r - \r^\prime) \rho(\r^\prime)
\end{equation}
We note that the kinetic term corresponding to quantum fluctuations is proportional to $J$, which accounts for the increase of $E_{2e}$ as a function of $J$ at fixed $\lambda$ that cannot be captured by numerics on the classical model. This is in excellent agreement with several non-trivial features of our DMRG results. First we note that the energy increase of a $2e$ skyrmion for fixed (small) $\lambda$ is linear in $J$. For a given $\lambda$, the size of the skyrmion remains fixed and the isospins from the two layers maintain local antiferromagnetism, implying that the classical energy is independent of $J$. Therefore, the correction to the NL$\sigma$M energy comes entirely from quantum fluctuations on the same classical texture, and therefore grows linearly with $J$. Next, we note that $E_{2e}$ for DMRG and NL$\sigma$M are very close when $\lambda$ is small, corresponding to large skyrmionic textures and small quantum corrections. As $\lambda$ grows larger, the $2e$ skyrmion wave-packet grows smaller in real space, and the kinetic energy contribution increases in accordance with the Heisenberg uncertainty principle $\Delta L \Delta n \gtrsim 1$. Accordingly, we show in Fig.~\ref{fig:En2eVsJ} that $E_{2e}$ increases with $J$ at a faster rate for larger $\lambda$. Similar considerations also apply to $E_{1e}$, which is generally higher in DMRG than in NL$\sigma$M --- however generally we do not expect the NL$\sigma$M estimates to be too accurate for charge $e$ excitations on top of the insulating state, as they are well-localized in real space.

\subsection{Particle-hole gap in superconductor}
\label{appss:PHgap}
\begin{figure}
    \centering
    \includegraphics[width =0.8\textwidth]{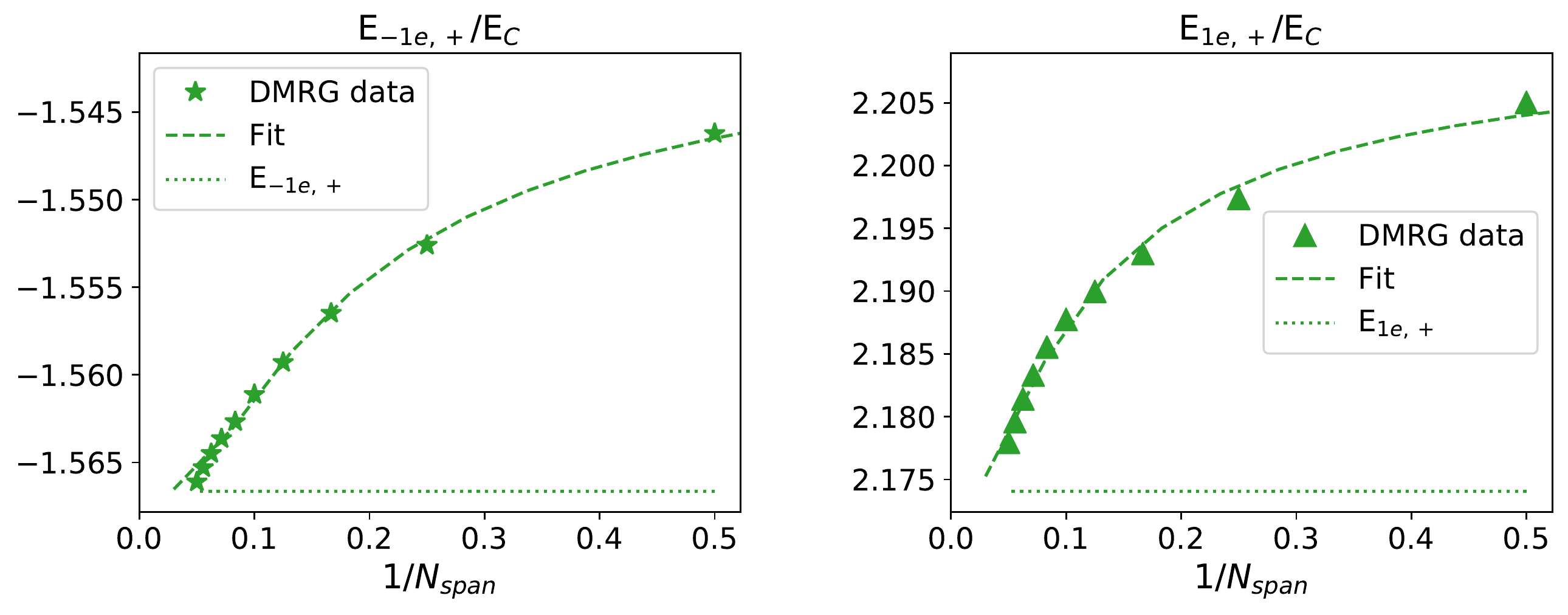}
    \caption{Segment DMRG data for energy particle and hole-like excitations in the top layer ($\gamma = +$) at $(J,\lambda) = (1.0,0.5)$ and $\nu = 2 + 1/4$, with best exponential fit and extrapolated value at $L_y = 10 \ell_B$ and $\chi = 2700$.  We find that a much larger $N_{span}$ is required for convergence indicating spatially extended electronic states, in contrast to the insulator at $\nu =2$. Slightly smaller values may be obtained by $L_y$ extrapolation, which we did not perform for this dataset.}
    \label{fig:dopedPHgapVsSpan}
\end{figure}

In order to show that the superconductor obtained in our DMRG study has a spectral gap to single-particle excitations, we perform segment DMRG in the doped phase. Specifically, we consider a representative point deep in the superconducting phase at $(J,\lambda) = (1.0,0.5)$ at two different fillings $\nu = 2 + 1/4$ and $\nu = 2 + 1/2$; and compute the energy required to add or remove an electron in a single layer (with given isospin). For $L_y = 10$ and $\chi = 2700$, we find that the sum of these energies, which corresponds to the particle-hole gap and is independent of where the chemical potential lies in the gap, is equal to $0.61 E_C$ ($0.55 E_C$) for $\nu - 2 = 1/4 $ ($1/2$), as shown in Fig.~\ref{fig:dopedPHgapVsSpan} for the smaller doping. Further, this gap is independent of the layer and spin quantum numbers of the particle; this is consistent with the superconductor being an isospin singlet. 

\section{$L_y$ dependence of $J_c$ \label{App:LyJc}}

\begin{figure}
    \centering
    \includegraphics[width=0.5\textwidth]{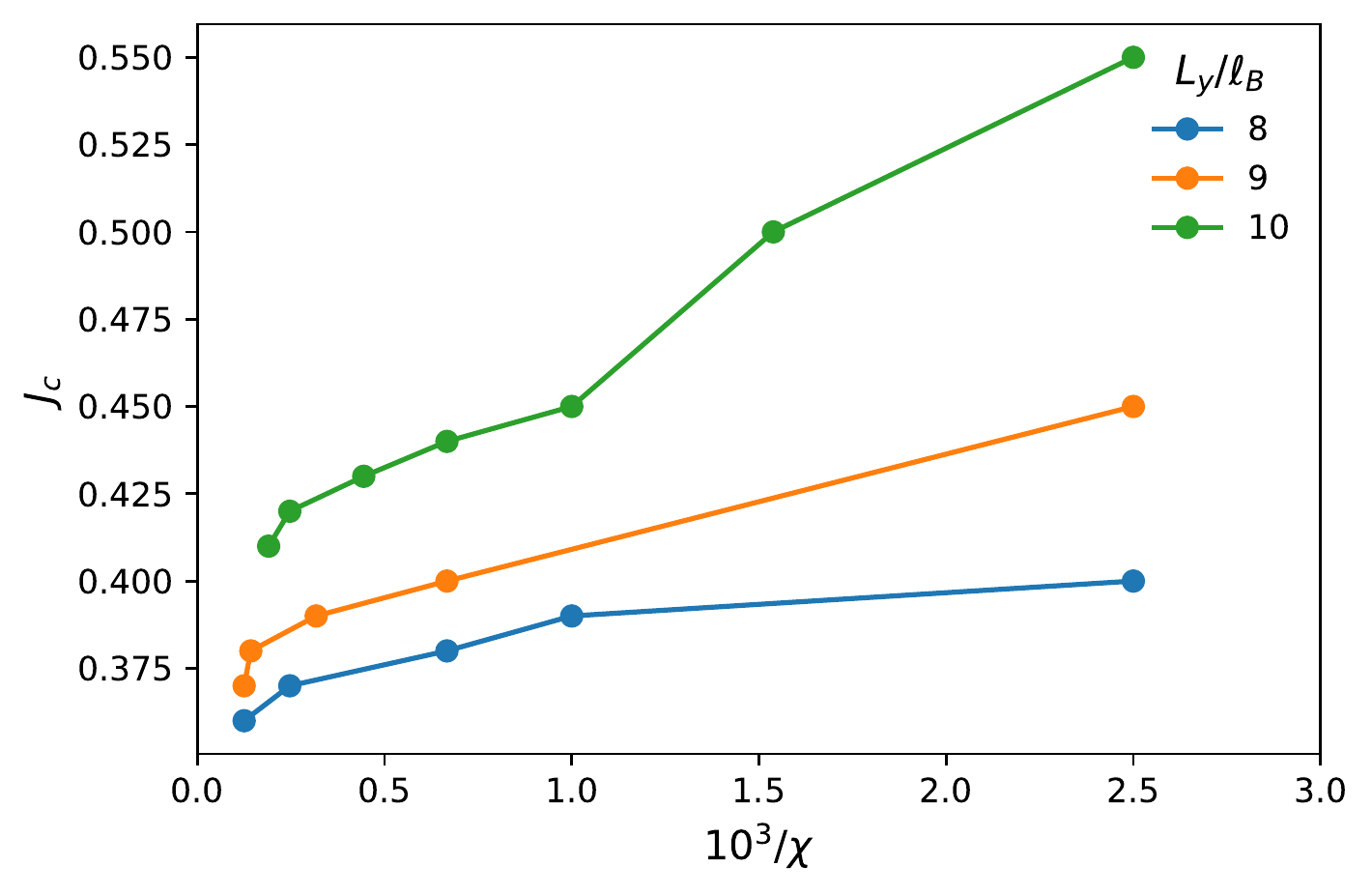}
    \caption{Dependence of the CDW/SC phase boundary $J_c$ on system size $L_y$ and DMRG bond dimension $\chi$, at $\lambda = 0.1$ and doping $\delta = 1/4$.}
    \label{fig:Jc_vs_chi}
\end{figure}

Here we explain why it is difficult to extrapolate the CDW/SC boundary $J_c(L_y)$ of the doped phase diagram to $L_y \to \infty$.
Recall that calculations are actually done at finite MPS bond dimension $\chi$, which (for small $\chi$) introduces a variational bias against strongly entangled states.
We find that the SC has much more entanglement than the CDW. This bias in favor of the CDW at finite $\chi$ results in the critical $J_c$ moving upward. This effect is confirmed in Fig.~\ref{fig:Jc_vs_chi}, where we show the behavior of $J_c(L_y, \chi)$ for various $L_y$. 
Unfortunately, since the entanglement scales in proportion to the circumference $L_y$, this finite-$\chi$ bias is more severe at larger $L_y$ (e.g. the curve for $L_y=10\ell_B$ is steeper than the one for $L_y=8\ell_B$).
Thus, without careful extrapolation in $\chi \to \infty$, one may spuriously conclude that $J_c$ increases with $L_y$, possibly indicating an instability of the SC phase in the planar limit $L_y\gg \ell_B$. 
As can be seen in Fig.~\ref{fig:Jc_vs_chi}, however, the critical coupling $J_c$ is still strongly drifting at our highest accessible bond dimension $\chi \simeq 8000$, so that a reliable extrapolation does not seem possible with our current data set.

\section{Pair wavefunction and the absence of higher-angular momentum pair correlations \label{app:pairwf} }
In the main text we show results for the pair order parameter $\Delta(\mathbf{r}) = i \eta^y_{ij} \psi_{+,i}(\mathbf{r}) \psi_{-,j}(\mathbf{r})$, which has zero orbital angular momentum. The charges in the pair can thus sit directly on top of each other, and in this sense the superconductor is ``$s$-wave.''
However, to investigate the possibility of higher-order pairing on the same footing, we can consider a generalized pairing function
\begin{align}
\Delta(\mathbf{r}, \mathbf{R}) = i \eta^y_{ij} e^{i \int^{\mathbf{r} + \mathbf{R}/2}_{\mathbf{r} - \mathbf{R}/2} \mathbf{A}.d\mathbf{R} } \psi_{+,i}(\mathbf{r} + \mathbf{R}/2) \psi_{-,j}(\mathbf{r} - \mathbf{R}/2)
\end{align}
The phase factor $\mathbf{A}.d\mathbf{R}$ is included to make the expression gauge invariant. In the symmetric gauge $\mathbf{A} = B(-y, x)/2$ centered on $\mathbf{r} = 0$, it vanishes, and the angular momentum of the pair is diagnosed by expanding in powers of $(R_x + i R_y)^m$ as usual. 
In the 2D limit, $\langle \Delta(0, \mathbf{R}) \rangle$ would measure the pairing wavefunction. 
Note that because of the inclusion of the layer index $\pm$, the pairing can in principle of have either even orbital angular momentum ($\Delta(0, \mathbf{R}) = \Delta(0, -\mathbf{R})$), or odd angular momentum ($\Delta(0, \mathbf{R}) = -\Delta(0, -\mathbf{R})$); in the odd case the pair is both an isospin and layer singlet. 

\begin{figure}[t]
    \includegraphics[width=0.3\columnwidth]{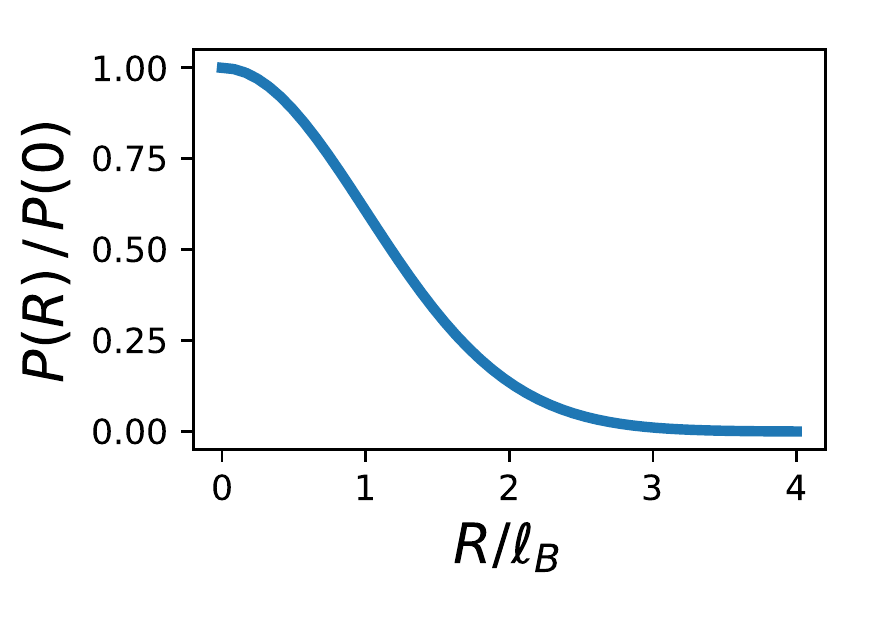}
    \caption{Pairing wavefunction in the superconductor of Fig 3c. \label{fig:Pr}}
\end{figure}

On the cylinder, where the order is algebraic, we can instead measure $P(\mathbf{R}) = \int d^2 r \langle \Delta^\dagger(\mathbf{r}, \mathbf{R})  \Delta(0, \mathbf{R}) \rangle$, which is like the pairing function squared. 
$P(\mathbf{R})$, shown in Fig.~\ref{fig:Pr}, has a maxima at $\mathbf{R} = 0$ consistent with $s$-wave pairing. 
Interpreting the width of $P(\mathbf{R})$ as the coherence length, we find $\xi \sim \ell_B$ (in fact $P(\mathbf{R}) = e^{-\frac{R^2}{2 \ell_B^2}}$).
This corresponds to an $L_M$-scale coherence length, as observed in MATBG experiments. 

\end{document}